\documentclass[english]{revtex4}
\usepackage[T1]{fontenc}
\usepackage[latin1]{inputenc}

\makeatletter

\usepackage{babel}
\makeatother
\begin{document}

\title{Regularization parameters for the self-force of a scalar particle
in a general orbit about a Schwarzschild black hole}

\author{Dong-Hoon Kim}

\address{Center for Quantum Spacetime, \#310 Ricci Annex Hall, Sogang University,
Shinsu-dong Mapogu Seoul 121-742, Korea}

\begin{abstract}
The interaction of a charged particle with its own field results in
the \char`\"{}self-force\char`\"{} on the particle, which includes
but is more general than the radiation reaction force. In the vicinity
of the particle in curved spacetime, one may follow Dirac and split
the retarded field of the particle into two parts, (1) the singular
source field, $\sim q/r$, and (2) the regular remainder field. The
singular source field exerts no force on the particle, and the self-force
is entirely caused by the regular remainder. We describe an elementary
multipole decomposition of the singular source field which allows
for the calculation of the self-force on a scalar-charged particle
orbiting a Schwarzschild black hole. 
\end{abstract}
\maketitle

\section{INTRODUCTION\label{sec:INTRODUCTION}}

According to the equivalence principle in general relativity, a particle
of infinitesimal mass orbits a black hole of large mass along a geodesic
worldline $\Gamma$ in the background spacetime determined by the
large mass alone. For a particle of small but finite mass, the orbit
is no longer a geodesic in the background of the large mass because
the particle perturbs the spacetime geometry.  This perturbation due
to the presence of the smaller mass modifies the orbit of the particle
from an original geodesic in the background. The difference of the
actual orbit from a geodesic in the background is said to result from
the interaction of the moving particle with its own gravitational
field, which is called a \emph{self-force} \citet{detweiler-whiting(03)}. 

Historically, Dirac \citet{dirac(38)} first gave the analysis of
the self-force for the electromagnetic field of a particle in flat
spacetime. He was able to approach the problem in a perturbative scheme
by allowing the particle's size to remain finite and invoking the
conservation of the stress-energy tensor inside a narrow world tube
surrounding the particle's worldline. Dewitt and Brehme \citet{dewitt-brehme(60)}
extended Dirac's problem to curved spacetime. Mino, Sasaki, and Tanaka
\citet{mino-sasaki-tanaka(97)} generalized it for the gravitational
field self-force. Quinn and Wald \citet{quinn-wald(97)} and Quinn
\citet{quinn(00)} worked out similar schemes for the gravitational,
electromagnetic, and scalar field self-forces by taking an axiomatic
approach. 

In Dirac's \citet{dirac(38)} flat spacetime problem, the retarded
field is decomposed into two parts: (\emph{i}) The first part is the
{}``mean of the advanced and retarded fields'' which is a solution
of the inhomogeneous field equation resembling the Coulomb $q/r$
piece of the scalar potential near the particle, (\emph{ii}) The second
part is a {}``radiation'' field which is a homogeneous solution
of Maxwell's equations. Dirac describes the self-force as the interaction
of the particle with the radiation field, a well-defined vacuum field
solution.

In the analyses of the self-force in curved spacetime, the Hadamard
form of Green's function \citet{dewitt-brehme(60)} is employed to
describe the retarded field of the particle. Traditionally, taking
the scalar field case for example, the retarded Green's function $G^{\mathrm{ret}}(p,p')$
is divided into {}``direct'' and {}``tail'' parts: (\emph{i})
The first part has support only on the past null cone of the field
point $p$, (\emph{ii}) The second part has the support inside the
past null cone due to the presence of the curvature of spacetime.
Accordingly, the self-force on the particle would consist of two pieces:
(\emph{i}) The first piece comes from the direct part of the field
and the acceleration of the worldline in the background geometry;
this corresponds to Abraham-Lorentz-Dirac (ALD) force in flat spacetime,
(\emph{ii}) The second piece comes from the tail part of the field
and is present in curved spacetime. Thus, the description of the self-force
in curved spacetime should reduce to Dirac's result in the flat spacetime
limit. In this approach, the self-force is considered to result via

\begin{equation}
\mathcal{F}_{a}=q\nabla_{a}\psi,\label{eq:1}\end{equation}
from the interaction of the particle with the quantity \citet{detweiler-whiting(03)}\begin{equation}
\psi^{\mathrm{\mathrm{sel}f}}=\psi^{\mathrm{ret}}-\psi^{\mathrm{dir}}.\label{eq:psi_self}\end{equation}

Although this traditional approach provides adequate methods to compute
the self-force, it does not share the physical simplicity of Dirac's
analysis where the force is described entirely in terms of an identifiable,
vacuum solution of the field equations: unlike Dirac's radiation field,
the quantity in Eq.~(\ref{psi_self}) is not a homogeneous solution
of the field equation $\nabla^{2}\psi=-4\pi\varrho$. Moreover, the
integral term in this quantity comes from the tail part of the Green's
function and is generally not differentiable on the worldline if the
Ricci scalar of the background is not zero (similarly, the electromagnetic
potential $A_{a}^{\mathrm{tail}}$ and the gravitational metric perturbation
$h_{ab}^{\mathrm{tail}}$ are not differentiable at the point of the
particle unless $\left(R_{ab}-\frac{{1}}{6}g_{ab}R\right)u^{b}$ and
$R_{cadb}u^{c}u^{d}$, respectively are zero in the background \citet{detweiler-m-w(03)}).
Thus, some version of averaging process must be invoked to make sense
of the self-force.

In this paper an alternative method to split the retarded field $\psi^{\mathrm{ret}}$
as suggested by Ref.~\citet{detweiler-whiting(03)} is used such
that the splits resemble those by Dirac: (\emph{i}) The first part,
the \emph{Singular Source field} $\psi^{\mathrm{S}}$, is an inhomogeneous
field similar to the Coulomb potential piece and exerts no force on
the particle, (\emph{ii}) The second part, the \emph{Regular Remainder
field} $\psi^{\mathrm{R}}$, being an homogeneous solution of the
field equation, is analogous to Dirac's radiation field and entirely
responsible for the self-force. This alternative method is reviewed
briefly in Section~\ref{sec:DECOMPOSITION-OF-THE}.

In Section~\ref{sec:MODE-SUM-DECOMPOSITION-AND} we give a brief
overview of the mode-sum decomposition scheme to evaluate the self-force.
We consider a particle with a scalar charge $q$ in general geodesic
motion about a Schwarzschild black hole in this paper. A spherical
harmonic decomposition of both $\psi^{\mathrm{ret}}$ and $\psi^{\mathrm{S}}$
having this condition should be performed to provide the multipole
components of each. Then, the mode by mode sum of the difference of
these components determines $\psi^{\mathrm{R}}$, and, thence the
self-force. The multipole components of $\psi^{\mathrm{ret}}$ can
be determined numerically while the multipole components of $\psi^{\mathrm{S}}$
are derived analytically. In particular, the multipole moments of
$\psi^{\mathrm{S}}$ are generically referred to as the \emph{Regularization
Parameters}; the entire paper focuses on the analytical task to find
these regularization parameters. We summarize our analytical results
at the end of the Section. 

The description of $\psi^{\mathrm{S}}$ becomes advantageously simple
in a specially chosen co-moving frame: the THZ normal coordinates
of this frame, introduced by Thorne and Hartle \citet{thorne-hartle(85)}
and extended by Zhang \citet{zhang(86)} are locally inertial on a
geodesic. In Section~\ref{sec:DETERMINATION-OF-psiS} we obtain a
simple form of $\psi^{\mathrm{S}}$ using the THZ coordinates and
then re-express it in terms of the background, i.e. Schwarzschild
coordinates via the coordinate transformation between the THZ and
the background coordinates.

Section~\ref{sec:-REGULARIZATION-PARAMETERS} outlines the derivation
of the regularization parameters given below in Eqs.~(\ref{eq:10})
to (\ref{eq:17}). These results are in agreement with Barack and
Ori \citet{barack-ori(02)} and Mino, Nakano, and Sasaki \citet{mino-nakano-sasaki(02)}. 

In Appendix \ref{app.A.THZ} we give a detailed description of the
THZ coordinates. Appendix \ref{hyper} provides some mathematical
details concerning the hypergeometric functions and the different
representations of the regularization parameters in connection with
them.\\
\textbf{}\\
\textbf{Notation:} $\left(t,\, r,\,\theta,\,\phi\right)$ are the
usual Schwarzschild coordinates. $\left(T,\, X,\, Y,\, Z\right)$
are the initial static normal coordinates, reshaped out of the Schwarzschild
coordinates. $\left(\mathcal{T},\,\mathcal{X},\,\mathcal{Y},\,\mathcal{Z}\right)$
are the THZ normal coordinates attached to the geodesic $\Gamma$,
and $\rho\equiv\sqrt{{\mathcal{X}^{2}+\mathcal{Y}^{2}+\mathcal{Z}^{2}}}$.
The points $p$ and $p'$ refer to a field point and a source point
on the worldline of the particle, respectively. In the coincidence
limit $p\rightarrow p'$.

\section{DECOMPOSITION OF THE RETARDED FIELD\label{sec:DECOMPOSITION-OF-THE}}

The recent analysis of the Green's function decomposition by Detweiler
and Whiting \citet{detweiler-whiting(03)} shows an alternative way
to split the retarded field into two parts 

\begin{equation}
\psi^{\mathrm{ret}}=\psi^{\mathrm{S}}+\psi^{\mathrm{R}},\label{eq:2}\end{equation}
where $\psi^{\mathrm{S}}$ and $\psi^{\mathrm{R}}$ are named the
\textbf{S}ingular \textbf{S}ource field and the \textbf{R}egular \textbf{R}emainder
field, respectively. The source function for a point particle on the
worldline $\Gamma$ is $\varrho(p)=q\int(-g)^{-1/2}\delta^{4}(p-p'(\tau'))d\tau'$.
Like $\psi^{\mathrm{ret}}$, $\psi^{\mathrm{S}}$ is an inhomogeneous
solution of the scalar field equation 

\begin{equation}
\nabla^{2}\psi=-4\pi\varrho\label{eq:3}\end{equation}
in the neighborhood of the particle. And $\psi^{\mathrm{S}}$ is determined
in the neighborhood of the particle's worldline entirely by local
analysis. $\psi^{\mathrm{R}}$, defined by Eq.~(\ref{eq:2}), is
then necessarily a homogeneous solution and is therefore expected
to be differentiable on $\Gamma$. According to Ref.~\citet{detweiler-whiting(03)},
$\psi^{\mathrm{R}}$ will formally give the correct self-force when
substituted on the right hand side of Eq.~(\ref{eq:1}) in place
of $\psi^{\mathrm{tail}}$. In this paper we adopt this decomposition,
and try to determine an analytical approximation to $\psi^{\mathrm{S}}$,
which is to be subtracted from $\psi^{\mathrm{ret}}$ for an explicit
computation of the self-force.

\section{MODE-SUM DECOMPOSITION AND REGULARIZATION PARAMETERS\label{sec:MODE-SUM-DECOMPOSITION-AND}}

By Eq.~(\ref{eq:1}) the self-force can be formally evaluated from\begin{eqnarray}
\mathcal{F}_{a}^{\mathrm{self}} & = & \lim_{p\rightarrow p'}\left[\mathcal{F}_{a}^{\mathrm{ret}}(p)-\mathcal{F}_{a}^{\mathrm{S}}(p)\right]=\lim_{p\rightarrow p'}\mathcal{F}_{a}^{\mathrm{R}}(p)\nonumber \\
 & \equiv & q\lim_{p\rightarrow p'}\nabla_{a}\psi^{\mathrm{R}}=q\lim_{p\rightarrow p'}\nabla_{a}(\psi^{\mathrm{ret}}-\psi^{\mathrm{S}}),\label{eq:4}\end{eqnarray}
where $p'$ is the event on $\Gamma$ where the self-force is to be
determined and $p$ is an event in the neighborhood of $p'$. For
use of this equation, both $\mathcal{F}_{a}^{\mathrm{ret}}(p)$ and
$\mathcal{F}_{a}^{\mathrm{S}}(p)$ would be expanded into multipole
$\ell$-modes, with $\mathcal{F}_{\ell a}^{\mathrm{ret}}(p)$ determined
numerically.

Typically, if the background geometry is Schwarzschild spacetime,
the source function $\varrho(p)$ is expanded in terms of spherical
harmonics, and then similar expansion for $\psi^{\mathrm{ret}}$ is
made

\begin{equation}
\psi^{\mathrm{ret}}={\displaystyle \sum_{\ell m}}\psi_{\ell m}^{\mathrm{ret}}(r,t)Y_{\ell m}(\theta,\phi),\label{eq:5}\end{equation}
 where $\psi_{\ell m}^{\mathrm{ret}}(r,t)$ is found numerically.
The individual $\ell m$ components of $\psi^{\mathrm{ret}}$ in this
expansion are finite at the location of the particle even though their
sum is singular. Then, $\mathcal{F}_{\ell a}^{\mathrm{ret}}$ is finite
and can be obtained as

\begin{equation}
\mathcal{F}_{\ell a}^{\mathrm{ret}}=q\nabla_{a}\sum_{m}\psi_{\ell m}^{\mathrm{ret}}Y_{\ell m},\label{eq:6}\end{equation}
where $a$ represents each component of $t$, $r$, $\phi$, $\theta$
in the Schwarzschild geometry.

The singular source field $\psi^{\mathrm{S}}$ is determined analytically
in the neighborhood of the particle's worldline via local analysis
(see Section \ref{sec:DETERMINATION-OF-psiS}). Then, $\nabla_{a}\psi^{\mathrm{S}}$
is evaluated and the mode-sum decomposition of this quantity is performed
to provide 

\begin{equation}
\mathcal{F}_{\ell a}^{\mathrm{S}}=q\nabla_{a}\sum_{m}\psi_{\ell m}^{\mathrm{S}}Y_{\ell m},\label{eq:7}\end{equation}
which is also finite at the location of the particle.

Then, using Eqs. (\ref{eq:4}), (\ref{eq:6}), and (\ref{eq:7}) the
self-force is finally\begin{eqnarray}
\mathcal{F}_{a}^{\mathrm{self}} & = & \sum_{\ell}\lim_{p\rightarrow p'}\left[\mathcal{F}_{\ell a}^{\mathrm{ret}}(p)-\mathcal{F}_{\ell a}^{\mathrm{S}}(p)\right]\nonumber \\
 & = & q\sum_{\ell}\lim_{p\rightarrow p'}\nabla_{a}\sum_{m}(\psi_{\ell m}^{\mathrm{ret}}-\psi_{\ell m}^{\mathrm{S}})Y_{\ell m}\label{eq:8}\end{eqnarray}
evaluated at the location of the particle.

In Section \ref{sec:-REGULARIZATION-PARAMETERS} the regularization
parameters are derived from the multipole components of $\nabla_{a}\psi^{\mathrm{S}}$
evaluated at the source point and are used to control both singular
behavior and differentiability. We follow Barack and Ori \citet{barack-ori(00)}
in defining the regularization counter terms, except that the singular
source field $\psi^{\mathrm{S}}$ is used in place of $\psi^{\mathrm{dir}}$
\begin{equation}
\lim_{p\rightarrow p'}\mathcal{F}_{\ell a}^{\mathrm{S}}=\left(\ell+\frac{{1}}{2}\right)A_{a}+B_{a}+\frac{{C_{a}}}{\ell+\frac{{1}}{2}}+O(\ell^{-2}),\label{eq:9}\end{equation}
and show 

\begin{equation}
A_{t}=\mathrm{sgn}(\Delta)\frac{{q^{2}}}{r_{\mathrm{o}}^{2}}\frac{{\dot{r}}}{1+J^{2}/r_{\mathrm{o}}^{2}},\label{eq:10}\end{equation}

\begin{equation}
A_{r}=-\mathrm{sgn}(\Delta)\frac{{q^{2}}}{r_{\mathrm{o}}^{2}}\frac{{E}\left(1-\frac{{2M}}{r_{\mathrm{o}}}\right)^{-1}}{1+J^{2}/r_{\mathrm{o}}^{2}},\label{eq:11}\end{equation}

\begin{equation}
A_{\phi}=0,\label{eq:12}\end{equation}

\begin{equation}
B_{t}=\frac{{q^{2}}}{r_{\mathrm{o}}^{2}}E\dot{r}\left[\frac{{F_{3/2}}}{\left(1+J^{2}/r_{\mathrm{o}}^{2}\right)^{3/2}}-\frac{{3F_{5/2}}}{2\left(1+J^{2}/r_{\mathrm{o}}^{2}\right)^{5/2}}\right],\label{eq:13}\end{equation}
\begin{equation}
B_{r}=\frac{{q^{2}}}{r_{\mathrm{o}}^{2}}\left\{ -\frac{{F_{1/2}}}{\left(1+J^{2}/r_{\mathrm{o}}^{2}\right)^{1/2}}+\frac{{[1-2\left(1-\frac{{2M}}{r_{\mathrm{o}}}\right)^{-1}\dot{{r}}^{2}]{F_{3/2}}}}{2\left(1+J^{2}/r_{\mathrm{o}}^{2}\right)^{3/2}}+\frac{{3}\left(1-\frac{{2M}}{r_{\mathrm{o}}}\right)^{-1}\dot{{r}}^{2}F_{5/2}}{2\left(1+J^{2}/r_{\mathrm{o}}^{2}\right)^{5/2}}\right\} ,\label{eq:14}\end{equation}

\begin{equation}
B_{\phi}=\frac{{q^{2}}}{J}\dot{r}\left[\frac{{F_{1/2}-F_{3/2}}}{\left(1+J^{2}/r_{\mathrm{o}}^{2}\right)^{1/2}}+\frac{{3(F_{5/2}-F_{3/2})}}{2\left(1+J^{2}/r_{\mathrm{o}}^{2}\right)^{3/2}}\right],\label{eq:15}\end{equation}

\begin{equation}
C_{t}=C_{r}=C_{\phi}=0,\label{eq:16}\end{equation}

\begin{equation}
A_{\theta}=B_{\theta}=C_{\theta}=0,\label{eq:17}\end{equation}
where $\Delta\equiv r-r_{\mathrm{o}}$, $E\equiv-u_{t}=\left(1-2M/r_{\mathrm{o}}\right)\left(dt/d\tau\right)_{\mathrm{o}}$
($\tau$: proper time) and $J\equiv u_{\phi}=r_{\mathrm{o}}^{2}\left(d\phi/d\tau\right)_{\mathrm{o}}$
are the conserved energy and angular momentum, respectively, and $\dot{r}\equiv u^{r}=\left(dr/d\tau\right)_{\mathrm{o}}$.
Also, shorthand notations are used for the hypergeometric functions,
$F_{p}\equiv{}_{2}F_{1}\left(p,\frac{{1}}{2};1;\frac{{J^{2}}}{r_{\mathrm{\mathrm{o}}}^{2}+J^{2}}\right)$
(see Appendix \ref{hyper} for more details about the hypergeometric
functions and the representations of the regularization parameters
in connection with them).

\section{DETERMINATION OF $\psi^{\mathrm{S}}$ VIA THE THZ NORMAL COORDINATES\label{sec:DETERMINATION-OF-psiS}}

It was mentioned earlier in Sections \ref{sec:DECOMPOSITION-OF-THE}
and \ref{sec:MODE-SUM-DECOMPOSITION-AND} that $\psi^{\mathrm{S}}$
is determined in the neighborhood of the particle's worldline entirely
by local analysis. When analyzed by some special local coordinate
system in which the background geometry looks as flat as possible,
the scalar wave equation might take a simple form and $\psi^{\mathrm{S}}$
might look like a simple Coulomb potential piece. Detweiler, Messaritaki,
and Whiting \citet{detweiler-m-w(03)} (cited henceforth as Paper
I) show 

\begin{equation}
\psi^{\mathrm{S}}=q/\rho+O(\rho^{2}/\mathcal{R}^{3}),\label{eq:18}\end{equation}
where $\rho=\sqrt{\mathcal{X}^{2}+\mathcal{Y}^{2}+\mathcal{Z}^{2}}$
with $\mathcal{X}$, $\mathcal{Y}$, $\mathcal{Z}$ being spatial
components in that local inertial coordinate system and $\mathcal{R}$
represents a length scale of the background geometry (the smallest
of the radius of curvature, the scale of inhomogeneities, and time
scale for changes in curvature along $\Gamma$).

Describing this special coordinate system more precisely, first, it
must be a \emph{normal} coordinate system where on $\Gamma$, the
metric and its first derivatives match the Minkowski metric, and the
coordinate $\mathcal{T}$ measures the proper time. Normal coordinates
for a geodesic, however, are not unique, and we use particular ones
that were introduced by Thorne and Hartle \citet{thorne-hartle(85)}
and extended by Zhang \citet{zhang(86)} to describe the external
multipole moments of a vacuum solution of the Einstein equations:
namely, the \textbf{THZ NORMAL COORDINATES}. 

However, in order to derive the regularization parameters from the
multipole components of $\nabla_{a}\psi^{\mathrm{S}}$, $\rho$ in
Eq.~(\ref{eq:18}) must be expressed in terms of the coordinates
of the original background, which is the Schwarzschild geometry in
our problem. Then, this requires us to find the expressions of $\mathcal{X}$,
$\mathcal{Y}$, $\mathcal{Z}$ (THZ) in terms of $t$, $r$, $\phi$,
$\theta$ (Schwarzschild): the task here is simply to find the coordinate
transformation between two different geometries. 

Based on the idea from Weinberg \citet{weinberg(72)}, one can achieve
this coordinate transformation to the level of accuracy we desire
for this particular problem of mode-sum regularization, by taking
the following two steps basically:

\begin{enumerate}
\item Find an inertial Cartesian coordinates $X^{A}$ to redirect the Schwarzschild
coordinates $x^{a}$ by the Taylor's expansion around the location
of the particle, $x_{\mathrm{o}}^{a}$; \begin{equation}
X^{A}=X_{\mathrm{o}}^{A}+M^{A}{}_{a}(x^{a}-x_{\mathrm{o}}^{a})+\frac{{1}}{2}M^{A}{}_{a}\left.\Gamma_{bc}^{a}\right|_{\mathrm{o}}(x^{b}-x_{\mathrm{o}}^{b})(x^{c}-x_{\mathrm{o}}^{c})+O[(x-x_{\mathrm{o}})^{3}],\label{eq:19}\end{equation}
 where we may choose $X_{\mathrm{o}}^{A}=0$ and $M^{A}{}_{a}=\mathrm{diag}\left[M^{T}{}_{t},\, M^{X}{}_{r},\, M^{Y}{}_{\phi},\, M^{Z}{}_{\theta}\right]$
for convenience (this choice will redirect and rescale the Schwarzschild
coordinates as $T=M^{T}{}_{t}(t-t_{\mathrm{o}})$, $X=M^{X}{}_{r}(r-r_{\mathrm{o}})$,
$Y=M^{Y}{}_{\phi}(\phi-\phi_{\mathrm{o}})$, $Z=M^{Z}{}_{\theta}(\theta-\theta_{\mathrm{o}})$.
\item Boost $X^{A}$ with $u^{A}$, the particle's four-velocity at $p'$
as measured in this Cartesian frame, to obtain the final coordinates
$\mathcal{X}^{A'}$; \begin{eqnarray}
\mathcal{X}^{A'} & = & \Lambda^{A'}{}_{A}X^{A}\nonumber \\
 & = & \Lambda^{A'}{}_{A}\left[M^{A}{}_{a}(x^{a}-x_{\mathrm{o}}^{a})+\frac{{1}}{2}M^{A}{}_{a}\left.\Gamma_{bc}^{a}\right|_{\mathrm{o}}(x^{b}-x_{\mathrm{o}}^{b})(x^{c}-x_{\mathrm{o}}^{c})\right]+O[(x-x_{\mathrm{o}})^{3}],\label{eq:20}\end{eqnarray}
where\begin{equation}
\Lambda^{A'}{}_{A}=\left[\begin{array}{cccc}
u^{T} & -u^{X} & -u^{Y} & -u^{Z}\\
 & 1+(u^{T}-1)(u^{X})^{2}/u^{2} & (u^{T}-1)u^{X}u^{Y}/u^{2} & (u^{T}-1)u^{X}u^{Z}/u^{2}\\
 & \textrm{SYM} & 1+(u^{T}-1)(u^{Y})^{2}/u^{2} & (u^{T}-1)u^{Y}u^{Z}/u^{2}\\
 &  &  & 1+(u^{T}-1)(u^{Z})^{2}/u^{2}\end{array}\right]\label{boost}\end{equation}
 with $u^{2}\equiv(u^{X})^{2}+(u^{Y})^{2}+(u^{Z})^{2}$ \citet{jackson(99)}.
\end{enumerate}
~~~According to Ref.~\citet{weinberg(72)}, one can show out of
Eq.~(\ref{eq:20})

\begin{eqnarray}
g^{A'B'} & = & g^{ab}\frac{{\partial\mathcal{X}^{A'}}}{\partial x^{a}}\frac{{\partial\mathcal{X}^{B'}}}{\partial x^{b}}\nonumber \\
 & = & \eta^{A'B'}+O[(x-x_{\mathrm{o}})^{2}],\; x^{a}\rightarrow x_{\mathrm{o}}^{a},\label{eq:21}\end{eqnarray}
so that\begin{equation}
\frac{{\partial g^{A'B'}}}{\partial\mathcal{X}^{C'}}=O[(x-x_{\mathrm{o}})],\; x^{a}\rightarrow x_{\mathrm{o}}^{a},\label{eq:22}\end{equation}
with the choice of $M^{T}{}_{t}=\left(1-\frac{{2M}}{r_{\mathrm{o}}}\right)^{1/2}$,
$M^{X}{}_{r}=\left(1-\frac{{2M}}{r_{\mathrm{o}}}\right)^{-1/2}$,
$M^{Y}{}_{\phi}=r_{\mathrm{o}}\sin\theta_{\mathrm{o}}$, $M^{Z}{}_{\theta}=-r_{\mathrm{o}}$.
Eqs.~(\ref{eq:21}) and (\ref{eq:22}) are the local inertial features
as expected for normal coordinates.

To simplify the calculations, we may confine the particle's orbits
to the equatorial plane $\theta_{\mathrm{o}}=\pi/2$. Then, we have\begin{equation}
M^{A}{}_{a}=\mathrm{diag}\left[f^{1/2},\, f^{-1/2},\, r_{\mathrm{o}},\,-r_{\mathrm{o}}\right],\label{eq:23}\end{equation}
 where $f\equiv\left(1-\frac{{2M}}{r_{\mathrm{o}}}\right)$. Also,
this constraint of the equatorial plane makes $u^{Z}=0$. We can rewrite
$u^{A}$ in terms of the Schwarzschild coordinates and the constants
of motion,\begin{equation}
u^{A}\equiv\left(u^{T},\, u^{X},\, u^{Y},\, u^{Z}\right)=\left(f^{-1/2}E,\, f^{-1/2}\dot{{r}},\,\frac{{J}}{r_{\mathrm{o}}},\,0\right),\label{eq:24}\end{equation}
where $E\equiv-u_{t}=f\left(dt/d\tau\right)_{\mathrm{o}}$ ($\tau$:
proper time) and $J\equiv u_{\phi}=r_{\mathrm{o}}^{2}\left(d\phi/d\tau\right)_{\mathrm{o}}$
are the conserved energy and angular momentum, respectively, and $\dot{r}\equiv u^{r}=\left(dr/d\tau\right)_{\mathrm{o}}$.
From this it follows that $u^{2}=f^{-1}E^{2}-1$ and we have

\begin{equation}
\Lambda^{A'}{}_{A}=\left[\begin{array}{cccc}
f^{-1/2}E & -f^{-1/2}\dot{{r}} & -J/r_{\mathrm{o}} & 0\\
 & 1+\dot{{r}}^{2}/(f^{1/2}E+f) & J\dot{{r}}/[r_{\mathrm{o}}(E+f^{1/2})] & 0\\
 & \textrm{SYM} & 1+J^{2}/[r_{\mathrm{o}}^{2}(f^{-1/2}E+1)] & 0\\
 &  &  & 1\end{array}\right].\label{eq:25}\end{equation}

Now we are finally able to express $\rho^{2}$ in terms of the Schwarzschild
coordinates. Using Eq.~(\ref{eq:20}) one may write\begin{eqnarray}
\rho^{2}=\mathcal{X}^{I}\mathcal{X}_{I} & = & \delta_{IJ}\Lambda^{I}{}_{C}\Lambda^{J}{}_{D}M^{C}{}_{c}M^{D}{}_{d}\left[(x^{c}-x_{\mathrm{o}}^{c})(x^{d}-x_{\mathrm{o}}^{d})+\left.\Gamma_{ab}^{c}\right|_{\mathrm{o}}(x^{a}-x_{\mathrm{o}}^{a})(x^{b}-x_{\mathrm{o}}^{b})(x^{d}-x_{\mathrm{o}}^{d})\right]\nonumber \\
 &  & +O[(x-x_{\mathrm{o}})^{4}],\label{eq:26}\end{eqnarray}
where $I,\, J=1,\,2,\,3$. Then, using Eqs.~(\ref{eq:23}) and (\ref{eq:25}),
Eq.~(\ref{eq:26}) can be eventually expressed as\begin{eqnarray}
\rho^{2} & = & (E^{2}-f)(t-t_{\mathrm{o}})^{2}-\frac{{2E\dot{{r}}}}{f}(t-t_{\mathrm{o}})(r-r_{\mathrm{o}})-2EJ(t-t_{\mathrm{o}})(\phi-\phi_{\mathrm{o}})\nonumber \\
 &  & +\frac{{1}}{f}\left(1+\frac{{\dot{{r}}^{2}}}{f}\right)(r-r_{\mathrm{o}})^{2}+\frac{{2J\dot{{r}}}}{f}(r-r_{\mathrm{o}})(\phi-\phi_{\mathrm{o}})+(r_{\mathrm{o}}^{2}+J^{2})(\phi-\phi_{\mathrm{o}})^{2}+r_{\mathrm{o}}^{2}\left(\theta-\frac{{\pi}}{2}\right)^{2}\nonumber \\
 &  & -\frac{{ME\dot{{r}}}}{r_{\mathrm{o}}^{2}}(t-t_{\mathrm{o}})^{3}+\frac{{M}}{r_{\mathrm{o}}^{2}}\left(-1+\frac{{2E^{2}}}{f}+\frac{{\dot{{r}}^{2}}}{f}\right)(t-t_{\mathrm{o}})^{2}(r-r_{\mathrm{o}})+\frac{{MJ\dot{{r}}}}{r_{\mathrm{o}}^{2}}(t-t_{\mathrm{o}})^{2}(\phi-\phi_{\mathrm{o}})\nonumber \\
 &  & -\frac{{ME\dot{{r}}}}{f^{2}r_{\mathrm{o}}^{2}}(t-t_{\mathrm{o}})(r-r_{\mathrm{o}})^{2}-\frac{{2(r_{\mathrm{o}}-M)EJ}}{fr_{\mathrm{o}}^{2}}(t-t_{\mathrm{o}})(r-r_{\mathrm{o}})(\phi-\phi_{\mathrm{o}})\nonumber \\
 &  & +r_{\mathrm{o}}E\dot{{r}}(t-t_{\mathrm{o}})(\phi-\phi_{\mathrm{o}})^{2}+r_{\mathrm{o}}E\dot{{r}}(t-t_{\mathrm{o}})\left(\theta-\frac{{\pi}}{2}\right)^{2}\nonumber \\
 &  & -\frac{{M}}{f^{2}r_{\mathrm{o}}^{2}}\left(1+\frac{{\dot{{r}}^{2}}}{f}\right)(r-r_{\mathrm{o}})^{3}+\frac{{(2r_{\mathrm{o}}-5M)J\dot{{r}}}}{f^{2}r_{\mathrm{o}}^{2}}(r-r_{\mathrm{o}})^{2}(\phi-\phi_{\mathrm{o}})\nonumber \\
 &  & +r_{\mathrm{o}}\left(1-\frac{{\dot{{r}}^{2}}}{f}+\frac{{2J^{2}}}{r_{\mathrm{o}}^{2}}\right)(r-r_{\mathrm{o}})(\phi-\phi_{\mathrm{o}})^{2}+r_{\mathrm{o}}\left(1-\frac{{\dot{{r}}^{2}}}{f}\right)(r-r_{\mathrm{o}})\left(\theta-\frac{{\pi}}{2}\right)^{2}\nonumber \\
 &  & -r_{\mathrm{o}}J\dot{{r}}(\phi-\phi_{\mathrm{o}})^{3}-r_{\mathrm{o}}J\dot{{r}}(\phi-\phi_{\mathrm{o}})\left(\theta-\frac{{\pi}}{2}\right)^{2}+O[(x-x_{\mathrm{o}})^{4}].\label{eq:27}\end{eqnarray}
Eq.~(\ref{eq:27}) is substituted into Eq.~(\ref{eq:18}) to determine
$\psi^{\mathrm{S}}$ in terms of the Schwarzschild coordinates and
will serve significantly to derive the regularization parameters in
the next section.

In the above analysis we have ignored the term $O[(x-x_{\mathrm{o}})^{3}]$
in Eq.~(\ref{eq:20}) and its contribution to $\rho^{2}$, which
is $O[(x-x_{\mathrm{o}})^{4}]$ in Eqs.~(\ref{eq:26}) and (\ref{eq:27}).
To the level of accuracy we desire for the mode-sum regularization
in this paper, that is to say, to the determination of \emph{C}-terms,
it is not necessary to specify the term $O[(x-x_{\mathrm{o}})^{4}]$
in $\rho^{2}$ (hence not necessary to specify $O[(x-x_{\mathrm{o}})^{3}]$
in the spatial THZ coordinates $\mathcal{X}$, $\mathcal{Y}$, $\mathcal{Z}$).
Even without specifying the term $O[(x-x_{\mathrm{o}})^{4}]$ in $\rho^{2}$,
one can prove that \emph{}$C_{a}$-terms in Eq.~(\ref{eq:9}) always
vanish (see Subsection \ref{sub:C-terms}). In fact, the clarification
of $O[(x-x_{\mathrm{o}})^{3}]$ for the THZ coordinates in Eq.~(\ref{eq:20})
requires more involved analyses of coordinate transformations, which
would be beyond the scope of this paper. Readers may refer to Appendix
\ref{app.A.THZ} for more detailed description of the THZ coordinates,
specified up to the quartic order. \\

\section{REGULARIZATION PARAMETERS FOR A GENERAL ORBIT OF THE SCHWARZSCHILD
GEOMETRY\label{sec:-REGULARIZATION-PARAMETERS}}

In Section \ref{sec:DETERMINATION-OF-psiS}, we have seen that an
approximation to $\psi^{\mathrm{S}}$ is\begin{equation}
\psi^{\mathrm{S}}=q/\rho+O(\rho^{2}/\mathcal{R}^{3}).\label{eq:28}\end{equation}
Following Paper I \citet{detweiler-m-w(03)}, the regularization parameters
can be determined from evaluating the multipole components of $\partial_{a}(q/\rho)$
($a=t,\, r,\,\theta,\,\phi$ for the Schwarzschild background) at
the location of the source. The error, $O(\rho^{2}/\mathcal{R}^{3})$
in the above approximation is disregarded since it gives no contribution
to $\nabla_{a}\psi^{\mathrm{S}}$ as we take the {}``coincidence
limit'', $x\rightarrow x_{\mathrm{o}}$ , where $x$ denotes a point
in the vicinity of the particle and $x_{\mathrm{o}}$ the location
of the particle in the Schwarzschild geometry. 

In evaluating the multipole components of $\partial_{a}(q/\rho)$,
singularities are expected with certain terms. To help identify those
singularities, we introduce an order parameter $\epsilon$ which is
to be set to unity at the end of a calculation: we attach $\epsilon^{n}$
to each $O[(x-x_{\mathrm{o}})^{n}]$ part of $\rho^{2}$ in Eq.~(\ref{eq:27})
and may re-express $\rho^{2}$ as\begin{equation}
\rho^{2}=\epsilon^{2}\mathcal{P}_{\mathrm{II}}+\epsilon^{3}\mathcal{P}_{\mathrm{III}}+\epsilon^{4}\mathcal{P}_{\mathrm{IV}}+O(\epsilon^{5}),\label{eq:29}\end{equation}
where $\mathcal{P}_{\mathrm{II}}$, $\mathcal{P}_{\mathrm{III}}$,
and $\mathcal{P}_{\mathrm{IV}}$ represent the quadratic, cubic, and
quartic order parts of $\rho^{2}$, respectively. Here we pretend
that the quartic part $\mathcal{P}_{\mathrm{IV}}$ is also specified:
this will help us to perform the structure analysis for $C_{a}$-terms
later in Subsection~\ref{sub:C-terms} when we prove that these regularization
parameters always vanish.

When we express $\partial_{a}\left(1/\rho\right)$ in the Laurent
series expansion to identify the terms according to their singular
patterns, every denominator of this expansion should take the form
of $\mathcal{P}_{\mathrm{II}}^{n/2}$ ($n=3,\,5,\,7$). Due to this
special position, which becomes singular in the coincidence limit,
$\mathcal{P}_{\mathrm{II}}$ would play a significant role in inducing
the multipole decomposition. 

However, the quadratic part $\mathcal{P}_{\mathrm{II}}$, directly
taken from Eq.~(\ref{eq:27}), may not be fully ready for this task
yet. First, $\phi-\phi_{\mathrm{o}}$ needs to be decoupled from $r-r_{\mathrm{o}}$
so that we have independent complete square forms of each, which is
necessary for inducing the Legendre polynomial expansions later. Coupling
between $t-t_{\mathrm{o}}$ and $\phi-\phi_{\mathrm{o}}$ is not significant
because upon fixing $t=t_{\mathrm{o}}$ all terms having $t-t_{\mathrm{o}}$
will vanish. Thus, we reshape our quadratic term in Eq.~(\ref{eq:27})
into\begin{eqnarray}
\mathcal{P}_{\mathrm{II}} & = & (E^{2}-f)(t-t_{\mathrm{o}})^{2}-\frac{{2E\dot{{r}}r_{\mathrm{o}}^{2}}}{f\left(r_{\mathrm{o}}^{2}+J^{2}\right)}(t-t_{\mathrm{o}})\Delta-2EJ(t-t_{\mathrm{o}})(\phi-\phi')\nonumber \\
 &  & +\frac{{E^{2}r_{\mathrm{o}}^{2}}}{f^{2}\left(r_{\mathrm{o}}^{2}+J^{2}\right)}\Delta^{2}+\left(r_{\mathrm{o}}^{2}+J^{2}\right)(\phi-\phi')^{2}+r_{\mathrm{o}}^{2}\left(\theta-\frac{{\pi}}{2}\right)^{2}\label{eq:33}\end{eqnarray}
with\begin{equation}
\phi'\equiv\phi_{\mathrm{o}}-\frac{{J\dot{{r}}\Delta}}{f\left(r_{\mathrm{o}}^{2}+J^{2}\right)},\label{eq:34}\end{equation}
where $\Delta\equiv r-r_{\mathrm{o}}$, and an identity $\dot{{r}}^{2}=E^{2}-f\left(1+J^{2}/r_{\mathrm{o}}^{2}\right)$
is used for simplifying the coefficient of $\Delta^{2}$. Here, taking
the coincidence limit $\Delta\rightarrow0$, we have $\phi'\rightarrow\phi_{\mathrm{o}}$
(the same idea is found in Mino, Nakano, and Sasaki~\citet{mino-nakano-sasaki(02)}).
Also, in order to be multipole-decomposed globally, the quadratic
part needs to be completely analytic and smooth over the entire two-sphere.
For this purpose we rewrite it as

\begin{eqnarray}
\mathcal{P}_{\mathrm{II}} & = & (E^{2}-f)(t-t_{\mathrm{o}})^{2}-\frac{{2E\dot{{r}}r_{\mathrm{o}}^{2}}}{f\left(r_{\mathrm{o}}^{2}+J^{2}\right)}(t-t_{\mathrm{o}})\Delta-2EJ(t-t_{\mathrm{o}})\sin\theta\sin(\phi-\phi')\nonumber \\
 &  & +\frac{{E^{2}r_{\mathrm{o}}^{2}}}{f^{2}\left(r_{\mathrm{o}}^{2}+J^{2}\right)}\Delta^{2}+(r_{\mathrm{o}}^{2}+J^{2})\sin^{2}\theta\sin^{2}(\phi-\phi')+r_{\mathrm{o}}^{2}\cos^{2}\theta\nonumber \\
 &  & +O[(x-x_{\mathrm{o}})^{4}],\label{qdrt}\end{eqnarray}
where one should notice that by replacing $\phi-\phi'=\sin(\phi-\phi')+O[(\phi-\phi')^{3}]$
and $1=\sin\theta+O[(\theta-\pi/2)^{2}]$ we create the $O[(x-x_{\mathrm{o}})^{4}]$
terms. 

To aid in the multipole decomposition we rotate the usual Schwarzschild
coordinates by following the approach of Barack and Ori~\citet{barack-ori(02)}
and Paper I \citet{detweiler-m-w(03)} such that the coordinate location
of the particle is moved from the equatorial plane $\theta=\frac{{\pi}}{2}$
to a location where $\sin\Theta=0$ ($\Theta$: new polar angle).
We define new angles $\Theta$ and $\Phi$ in terms of the usual Schwarzschild
angles by\begin{eqnarray}
\sin\theta\cos(\phi-\phi') & = & \cos\Theta\nonumber \\
\sin\theta\sin(\phi-\phi') & = & \sin\Theta\cos\Phi\nonumber \\
\cos\theta & = & \sin\Theta\sin\Phi.\label{eq:35}\end{eqnarray}
Also, under this coordinate rotation, a spherical harmonic $Y_{\ell m}(\theta,\phi)$
becomes\begin{equation}
Y_{\ell m}(\theta,\phi)=\sum_{m'=-\ell}^{\ell}\alpha_{mm'}^{\ell}Y_{\ell m'}(\Theta,\Phi),\label{eq:36}\end{equation}
where the coefficients $\alpha_{mm'}^{\ell}$ depend on the rotation
$(\theta,\phi)\rightarrow(\Theta,\Phi)$ as well as on $\ell$, $m$,
and $m'$, and the index $\ell$ is preserved under the rotation \citet{mathews-walker(70)}.
As recognized already in Paper I \citet{detweiler-m-w(03)}, there
is a great advantage of using the rotated angles $(\Theta,\Phi)$:
after expanding $\partial_{a}(q/\rho)$ into a sum of spherical harmonic
components, we take the coincidence limit $\Delta\rightarrow0$, $\Theta\rightarrow0$.
Then, finally only the $m=0$ components contribute to the self-force
at $\Theta=0$ since $Y_{\ell m}(0,\Phi)=0$ for $m\neq0$. Thus,
the regularization parameters of Eq.~(\ref{eq:9}) are just $(\ell,\, m=0)$
spherical harmonic components of $\partial_{a}(q/\rho)$ evaluated
at $x_{\mathrm{o}}^{a}$. 

Now, using these rotated angles, we may re-express Eq.~(\ref{qdrt})
as \begin{eqnarray}
\mathcal{P}_{\mathrm{II}} & = & (E^{2}-f)(t-t_{\mathrm{o}})^{2}-\frac{{2E\dot{{r}}r_{\mathrm{o}}^{2}}}{f\left(r_{\mathrm{o}}^{2}+J^{2}\right)}(t-t_{\mathrm{o}})\Delta-2EJ(t-t_{\mathrm{o}})\sin\Theta\cos\Phi\nonumber \\
 &  & +2\left(r_{\mathrm{o}}^{2}+J^{2}\right)\left(1-\frac{{J^{2}\sin^{2}\Phi}}{r_{\mathrm{o}}^{2}+J^{2}}\right)\left[\frac{{r_{\mathrm{o}}^{2}E^{2}\Delta^{2}}}{2f^{2}\left(r_{\mathrm{o}}^{2}+J^{2}\right)^{2}\left(1-\frac{{J^{2}\sin^{2}\Phi}}{r_{\mathrm{o}}^{2}+J^{2}}\right)}+1-\cos\Theta\right]\nonumber \\
 &  & +O[(x-x_{\mathrm{o}})^{4}],\label{qdrt3}\end{eqnarray}
where an approximation $\sin^{2}\Theta=2(1-\cos\Theta)+O(\Theta^{4})$
is used, the error from which is essentially $O(\Theta^{4})=O[(x-x_{\mathrm{o}})^{4}]$
and can be absorbed into $\mathcal{P}_{\mathrm{IV}}$. Here one should
note that through a series of modifications of the quadratic part
of Eq.~(\ref{eq:27}) we have created additional quartic order terms
apart from the desired form. Then, we may remove these additional
terms from the quadratic part and incorporate them into the quartic
part $\mathcal{P}_{\mathrm{IV}}$ in Eq.~(\ref{eq:29}). It is not
necessary, however, to specify this quartic part: as already mentioned
above, later in Subsection~\ref{sub:C-terms} we will show that the
quartic part $\mathcal{P}_{\mathrm{IV}}$ starts appearing from the
$\epsilon^{0}$-term and verify that the regularization parameters
of the $\epsilon^{0}$-term always vanish by analyzing the generic
structure of the $\epsilon^{0}$-term.

After removing $O[(x-x_{\mathrm{o}})^{4}]$ from Eq.~(\ref{qdrt3})
we may define\begin{eqnarray}
\tilde{{\rho}}^{2} & \equiv & (E^{2}-f)(t-t_{\mathrm{o}})^{2}-\frac{{2E\dot{{r}}r_{\mathrm{o}}^{2}}}{f\left(r_{\mathrm{o}}^{2}+J^{2}\right)}(t-t_{\mathrm{o}})\Delta-2EJ(t-t_{\mathrm{o}})\sin\Theta\cos\Phi\nonumber \\
 &  & +2\left(r_{\mathrm{o}}^{2}+J^{2}\right)\left(1-\frac{{J^{2}\sin^{2}\Phi}}{r_{\mathrm{o}}^{2}+J^{2}}\right)\left[\frac{{r_{\mathrm{o}}^{2}E^{2}\Delta^{2}}}{2f^{2}\left(r_{\mathrm{o}}^{2}+J^{2}\right)^{2}\left(1-\frac{{J^{2}\sin^{2}\Phi}}{r_{\mathrm{o}}^{2}+J^{2}}\right)}+1-\cos\Theta\right].\label{A}\end{eqnarray}
In particular, when fixing $t=t_{\mathrm{o}}$, we define \begin{equation}
\tilde{{\rho}}_{\mathrm{o}}^{2}\equiv\left.\tilde{{\rho}}^{2}\right|_{t=t_{\mathrm{o}}}=2\left(r_{\mathrm{o}}^{2}+J^{2}\right)\chi\left(\delta^{2}+1-\cos\Theta\right)\label{eq:37}\end{equation}
with

\begin{equation}
\chi\equiv1-\frac{{J^{2}\sin^{2}\Phi}}{r_{\mathrm{o}}^{2}+J^{2}}\label{eq:38}\end{equation}
and\begin{equation}
\delta^{2}\equiv\frac{{r_{\mathrm{o}}^{2}E^{2}\Delta^{2}}}{2f^{2}\left(r_{\mathrm{o}}^{2}+J^{2}\right)^{2}\chi}.\label{eq:39}\end{equation}

Now we rewrite Eq.~(\ref{eq:29}) by replacing the original quadratic
part $\mathcal{P}_{\mathrm{II}}$ with the modified form $\tilde{{\rho}}^{2}$
above, \begin{equation}
\rho^{2}=\epsilon^{2}\tilde{{\rho}}^{2}+\epsilon^{3}\mathcal{P}_{\mathrm{III}}+\epsilon^{4}\mathcal{P}_{\mathrm{IV}}+O(\epsilon^{5}),\label{new_rho^2}\end{equation}
where the new quartic part $\mathcal{P}_{\mathrm{IV}}$ includes the
additional quartic order terms that result from modification of the
quadratic part $\mathcal{P}_{\mathrm{II}}$. Then, based on this redefined
$\rho^{2}$, we have the following expression of $\left.\partial_{a}(1/\rho)\right|_{t=t_{\mathrm{o}}}$
in a Laurent series expansion \begin{equation}
\left.\partial_{a}\left(\frac{{1}}{\rho}\right)\right|_{t=t_{\mathrm{o}}}=-\frac{{1}}{2}\frac{{\left.\partial_{a}\left(\tilde{{\rho}}^{2}\right)\right|_{t=t_{\mathrm{o}}}}}{\tilde{{\rho}}_{\mathrm{o}}^{3}}\epsilon^{-2}+\left\{ -\frac{{1}}{2}\frac{{\left.\partial_{a}\mathcal{P}_{\mathrm{III}}\right|_{t=t_{\mathrm{o}}}}}{\tilde{{\rho}}_{\mathrm{o}}^{3}}+\frac{{3}}{4}\frac{{\left.\left[\partial_{a}\left(\tilde{{\rho}}^{2}\right)\right]\mathcal{P}_{\mathrm{III}}\right|_{t=t_{\mathrm{o}}}}}{\tilde{{\rho}}_{\mathrm{o}}^{5}}\right\} \epsilon^{-1}+O(\epsilon^{0}).\label{eq:30}\end{equation}

Eq.~(\ref{eq:37}), when inserted into Eq.~(\ref{eq:30}), plays
a very important role in calculating the regularization parameters
for the rest of the section: out of Eq.~(\ref{eq:37}), we induce
Legendre polynomial expansions in terms of $\cos\Theta$. Sometimes
we may have the dependence on $\tilde{{\rho}}_{\mathrm{o}}^{2}$ not
only in the denominators but also in the numerators on the right hand
side of Eq.~(\ref{eq:30}). In the numerators the dependence can
be found from the terms containing $\sin^{n}\Theta$ or $\cos^{n}\Theta$
since Eq.~(\ref{eq:37}) can be solved for $\cos\Theta$. After finding
all of the $\tilde{{\rho}}_{\mathrm{o}}^{2}$ dependence, the rest
of the calculations involve integrating over the angle $\Phi$ . The
techniques involved in Legendre polynomial expansions and integration
over $\Phi$ are described in detail in Appendices C and D of Paper
I \citet{detweiler-m-w(03)}.

Below in Subsections~\ref{sub:A-terms} and \ref{sub:B-terms}, we
present the key steps of calculating the regularization parameters.

\subsection{$A_{a}$-terms\label{sub:A-terms}}

We take the $\epsilon^{-2}$ term from Eq.~(\ref{eq:30}) and define\begin{equation}
Q_{a}[\epsilon^{-2}]\equiv-\frac{{q^{2}}}{2}\frac{{\left.\partial_{a}\left(\tilde{{\rho}}^{2}\right)\right|_{t=t_{\mathrm{o}}}}}{\tilde{{\rho}}_{\mathrm{o}}^{3}}\label{eq:40}\end{equation}
Then, we proceed with our calculations of the regularization parameters
one component at a time.

\subsubsection{$A_{t}$-term\emph{:}}

First \textbf{}we complete the expression for \textbf{$Q_{t}[\epsilon^{-2}]$}
by recalling \textbf{}Eqs.~(\ref{A}) and (\ref{eq:37})\textbf{\begin{eqnarray}
Q_{t}[\epsilon^{-2}] & = & -\frac{{q^{2}}}{2}\tilde{{\rho}}_{\mathrm{o}}^{-3}\left.\partial_{t}\left(\tilde{{\rho}}^{2}\right)\right|_{t=t_{\mathrm{o}}}\nonumber \\
 & = & \frac{{q^{2}}}{2}\left[2\left(r_{\mathrm{o}}^{2}+J^{2}\right)\chi\left(\delta^{2}+1-\cos\Theta\right)\right]^{-3/2}\left(\frac{{2E\dot{{r}}r_{\mathrm{o}}^{2}\Delta}}{f\left(r_{\mathrm{o}}^{2}+J^{2}\right)}+2EJ\sin\Theta\cos\Phi\right)\nonumber \\
 & = & \frac{{q^{2}E\dot{{r}}r_{\mathrm{o}}^{2}\Delta\chi^{-3/2}}}{2\sqrt{2}f\left(r_{\mathrm{o}}^{2}+J^{2}\right)^{5/2}}\left(\delta^{2}+1-\cos\Theta\right)^{-3/2}\nonumber \\
 &  & -\frac{{q^{2}EJ\chi^{-3/2}\cos\Phi}}{\sqrt{2}\left(r_{\mathrm{o}}^{2}+J^{2}\right)^{3/2}}\left.\frac{{\partial}}{\partial\Theta}\right|_{\Delta}\left(\delta^{2}+1-\cos\Theta\right)^{-1/2},\label{eq:41}\end{eqnarray}
}where $\left.\frac{{\partial}}{\partial\Theta}\right|_{\Delta}$
means that $\Delta$ is held constant while the differntiation is
performed with respect to $\Theta$.

According to Appendix D of Paper I \citet{detweiler-m-w(03)}, for
$p\geq1$ \begin{equation}
\left(\delta^{2}+1-\cos\Theta\right)^{-p-1/2}=\sum_{\ell=0}^{\infty}\frac{{2\ell+1}}{\delta^{2p-1}(2p-1)}\left[1+O(\ell\delta)\right]P_{\ell}(\cos\Theta),\;\delta\rightarrow0,\label{eq:42}\end{equation}
and for $p=0$\begin{equation}
\left(\delta^{2}+1-\cos\Theta\right)^{-1/2}=\sum_{\ell=0}^{\infty}\left[\sqrt{2}+O(\ell\delta)\right]P_{\ell}(\cos\Theta),\;\delta\rightarrow0.\label{rho^-1/2}\end{equation}
Then, by Eqs.~(\ref{eq:42}) for $p=1$, (\ref{rho^-1/2}), and (\ref{eq:39}),
in the limit $\delta\rightarrow0$ (equivalently $\Delta\rightarrow0$)
Eq.~(\ref{eq:41}) becomes 

\begin{eqnarray}
\lim_{\Delta\rightarrow0}Q_{t}[\epsilon^{-2}] & = & \mathrm{sgn}(\Delta)\frac{{q^{2}\dot{{r}}r_{\mathrm{o}}\chi^{-1}}}{\left(r_{\mathrm{o}}^{2}+J^{2}\right)^{3/2}}\sum_{\ell=0}^{\infty}\left(\ell+\frac{{1}}{2}\right)P_{\ell}(\cos\Theta)\nonumber \\
 &  & -\frac{{q^{2}EJ\chi^{-3/2}\cos\Phi}}{\left(r_{\mathrm{o}}^{2}+J^{2}\right)^{3/2}}\sum_{\ell=0}^{\infty}\left.\frac{{\partial}}{\partial\Theta}\right|_{\Delta}P_{\ell}(\cos\Theta).\label{eq:43}\end{eqnarray}

Then, we integrate $\lim_{\Delta\rightarrow0}Q_{t}[\epsilon^{-2}]$
over $\Phi$ and divide it by $2\pi$ (we denote this process by the
angle brackets {}``$\langle\,\rangle$'')\begin{equation}
\left\langle \lim_{\Delta\rightarrow0}Q_{t}[\epsilon^{-2}]\right\rangle =\mathrm{sgn}(\Delta)\frac{{q^{2}\dot{{r}}r_{\mathrm{o}}\left\langle \chi^{-1}\right\rangle }}{\left(r_{\mathrm{o}}^{2}+J^{2}\right)^{3/2}}\sum_{\ell=0}^{\infty}\left(\ell+\frac{{1}}{2}\right)P_{\ell}(\cos\Theta),\label{eq:44}\end{equation}
where we exploit the fact that $\left\langle \chi^{-3/2}\cos\Phi\right\rangle =0$
to get rid of the second part in Eq.~(\ref{eq:43}) %
\footnote{Or alternatively, one can use the arguement $\left.\frac{{\partial}}{\partial\Theta}\right|_{\Delta}P_{\ell}(\cos\Theta)=0$
as $\Theta\rightarrow0$, to show that this part does not survive
at the end.%
}. Appendix C of Paper I \citet{detweiler-m-w(03)} provides $\left\langle \chi^{-1}\right\rangle ={}_{2}F_{1}\left(1,\frac{{1}}{2};1;\alpha\right)\equiv F_{1}=\left(1-\alpha\right)^{-1/2}$,
where $\alpha\equiv J^{2}/\left(r_{\mathrm{o}}^{2}+J^{2}\right)$.
Substituting this into Eq.~(\ref{eq:44}), the regularization parameter
$A_{t}$ can be finally determined when we take the coefficient of
the sum on the right hand side in the coincidence limit $\Theta\rightarrow0$
\begin{equation}
A_{t}=\mathrm{sgn}(\Delta)\frac{{q^{2}}}{r_{\mathrm{o}}^{2}}\frac{{\dot{{r}}}}{1+J^{2}/r_{\mathrm{o}}^{2}}.\label{eq:45}\end{equation}

\subsubsection{$A_{r}$-term\emph{:}}

Similarly, we have\begin{equation}
Q_{r}[\epsilon^{-2}]=-\frac{{q^{2}}}{2}\tilde{{\rho}}_{\mathrm{o}}^{-3}\left.\partial_{r}\left(\tilde{{\rho}}^{2}\right)\right|_{t=t_{\mathrm{o}}}.\label{eq:46}\end{equation}
Here, before computing $\left.\partial_{r}\left(\tilde{{\rho}}^{2}\right)\right|_{t=t_{\mathrm{o}}}$
we reverse the process via Eqs.~(\ref{eq:33}), (\ref{qdrt}), (\ref{qdrt3}),
and (\ref{A}) to obtain the relation

\begin{eqnarray}
\tilde{\rho}^{2} & = & \mathcal{P}_{\mathrm{II}}+O[(x-x_{\mathrm{o}})^{4}]\nonumber \\
 & = & (E^{2}-f)(t-t_{\mathrm{o}})^{2}-\frac{{2E\dot{{r}}r_{\mathrm{o}}^{2}}}{f\left(r_{\mathrm{o}}^{2}+J^{2}\right)}(t-t_{\mathrm{o}})\Delta-2EJ(t-t_{\mathrm{o}})(\phi-\phi')\\
 &  & +\frac{{E^{2}r_{\mathrm{o}}^{2}}}{f^{2}\left(r_{\mathrm{o}}^{2}+J^{2}\right)}\Delta^{2}+\left(r_{\mathrm{o}}^{2}+J^{2}\right)(\phi-\phi')^{2}+r_{\mathrm{o}}^{2}\left(\theta-\frac{{\pi}}{2}\right)^{2}\nonumber \\
 &  & +O[(x-x_{\mathrm{o}})^{4}].\label{A-qdrt}\end{eqnarray}
Differentiating this with respect to $r$ and going through the process
via Eqs.~(\ref{qdrt}) and (\ref{eq:35}), Eq.~(\ref{eq:46}) can
be expressed with the help of Eq.~(\ref{eq:37}) as \begin{equation}
Q_{r}[\epsilon^{-2}]=-\frac{{q^{2}}}{f^{2}}\left[2\left(r_{\mathrm{o}}^{2}+J^{2}\right)\chi\left(\delta^{2}+1-\cos\Theta\right)\right]^{-3/2}\left[\frac{{r_{\mathrm{o}}^{2}E^{2}\Delta}}{r_{\mathrm{o}}^{2}+J^{2}}+fJ\dot{r}\sin\Theta\cos\Phi\right]\label{eq:Q_r}\end{equation}
\footnote{The by-product from $O[(x-x_{\mathrm{o}})^{4}]$ in Eq.~(\ref{A-qdrt}),
that is to say, $-\frac{{q^{2}}}{2}\tilde{{\rho}}_{\mathrm{o}}^{-3}O[(x-x_{\mathrm{o}})^{3}]$
is suppressed since this can be categorized into the $C_{a}$-term
group (see Subsection \ref{sub:C-terms}). %
}. Then, the rest of the calculation is carried out in the same fashion
as for the case of $A_{t}$-term above. We obtain\begin{equation}
A_{r}=-\mathrm{sgn}(\Delta)\frac{{q^{2}}}{r_{\mathrm{o}}^{2}}\frac{{E}}{f\left(1+J^{2}/r_{\mathrm{o}}^{2}\right)}.\label{eq:47}\end{equation}

\subsubsection{$A_{\phi}$-term\emph{:}}

First we have

\begin{eqnarray}
Q_{\phi}[\epsilon^{-2}] & = & -\frac{{q^{2}}}{2}\tilde{{\rho}}_{\mathrm{o}}^{-3}\left.\partial_{\phi}\left(\tilde{{\rho}}^{2}\right)\right|_{t=t_{\mathrm{o}}}.\label{eq:48}\end{eqnarray}
Taking the same steps as used for $A_{r}$-term above via Eqs.~(\ref{A-qdrt}),
(\ref{qdrt}), and (\ref{eq:35}) in order, we obtain\begin{equation}
\left.\partial_{\phi}\left(\tilde{{\rho}}^{2}\right)\right|_{t=t_{\mathrm{o}}}=2\left(r_{\mathrm{o}}^{2}+J^{2}\right)\sin\Theta\cos\Phi+O[(x-x_{\mathrm{o}})^{3}].\label{dA/dphi}\end{equation}
Then, in a similar manner to that employed in the previous cases,
in the limit $\Delta\rightarrow0$ Eq.~(\ref{eq:48}) becomes \begin{equation}
\lim_{\Delta\rightarrow0}Q_{\phi}[\epsilon^{-2}]=-\frac{{q^{2}\chi^{-3/2}\cos\Phi}}{\left(r_{\mathrm{o}}^{2}+J^{2}\right)^{1/2}}\sum_{\ell=0}^{\infty}\left.\frac{{\partial}}{\partial\Theta}\right|_{\Delta}P_{\ell}(\cos\Theta)\label{Q_phi}\end{equation}
\footnote{The by-product from $O[(x-x_{\mathrm{o}})^{3}]$ in Eq.~(\ref{dA/dphi}),
that is to say, $-\frac{{q^{2}}}{2}\tilde{{\rho}}_{\mathrm{o}}^{-3}O[(x-x_{\mathrm{o}})^{3}]$
is suppressed since this can be categorized into the $C_{a}$-term
group (see Subsection \ref{sub:C-terms}). %
}. The right hand side vanishes through {}``$\left\langle \,\right\rangle $''
process because $\left\langle \chi^{-3/2}\cos\Phi\right\rangle =0$.
Hence,\begin{equation}
A_{\phi}=0.\label{eq:49}\end{equation}

\subsubsection{$A_{\theta}$-term\emph{:} }

It is evident from the particle's motion, which is confined to the
equatorial plane $\theta_{\mathrm{o}}=\frac{{\pi}}{2}$, that no self
force is acting on the particle in the direction perpendicular to
this plane. This is due to the fact that both the derivatives of retarded
field and the singular source field with respect to $\theta$ tend
to zero in the coincidence limit. Our calculation of $A_{\theta}$
should support this. Through the same process as employed before,
we have\begin{equation}
Q_{\theta}[\epsilon^{-2}]=-\frac{{q^{2}}}{2}\tilde{{\rho}}_{\mathrm{o}}^{-3}\left.\partial_{\theta}\left(\tilde{{\rho}}^{2}\right)\right|_{t=t_{\mathrm{o}}}\label{eq:50}\end{equation}
with\begin{equation}
\left.\partial_{\theta}\left(\tilde{{\rho}}^{2}\right)\right|_{t=t_{\mathrm{o}}}=2r_{\mathrm{o}}^{2}\sin\Theta\sin\Phi+O[(x-x_{\mathrm{o}})^{3}].\label{dA/dtheta}\end{equation}
Then, similarly as in the case of $A_{\phi}$-term above\begin{equation}
\lim_{\Delta\rightarrow0}Q_{\theta}[\epsilon^{-2}]=-\frac{{q^{2}r_{\mathrm{o}}^{2}\chi^{-3/2}\sin\Phi}}{\left(r_{\mathrm{o}}^{2}+J^{2}\right)^{3/2}}\sum_{\ell=0}^{\infty}\left.\frac{{\partial}}{\partial\Theta}\right|_{\Delta}P_{\ell}(\cos\Theta).\label{Q_theta}\end{equation}
Again, via {}``$\left\langle \,\right\rangle $'' process, the right
hand side vanishes because $\left\langle \chi^{-3/2}\sin\Phi\right\rangle =0$.
Thus,\begin{equation}
A_{\theta}=0.\label{eq:51}\end{equation}

\subsection{$B_{a}$-terms\label{sub:B-terms}}

We take the $\epsilon^{-1}$ term from Eq.~(\ref{eq:30}) and define

\begin{equation}
Q_{a}[\epsilon^{-1}]\equiv q^{2}\left\{ -\frac{{1}}{2}\frac{{\left.\partial_{a}\mathcal{P}_{\mathrm{III}}\right|_{t=t_{\mathrm{o}}}}}{\tilde{{\rho}}_{\mathrm{o}}^{3}}+\frac{{3}}{4}\frac{{\left.\left[\partial_{a}\left(\tilde{{\rho}}^{2}\right)\right]\mathcal{P}_{\mathrm{III}}\right|_{t=t_{\mathrm{o}}}}}{\tilde{{\rho}}_{\mathrm{o}}^{5}}\right\} ,\label{eq:52}\end{equation}
where for computing $\partial_{a}\left(\tilde{{\rho}}^{2}\right)$,
Eq.~(\ref{A-qdrt}) should be referred to, and $\mathcal{P}_{\mathrm{III}}$
is the cubic part taken directly from Eq.~(\ref{eq:27}). 

We may express this in a generic form\begin{equation}
Q_{a}[\epsilon^{-1}]=\sum_{n=1}^{2}\sum_{k=0}^{2n}\sum_{p=0}^{[k/2]}\frac{{b_{nkp(a)}\Delta^{2n-k}\left(\phi-\phi_{\mathrm{o}}\right)^{k-2p}\left(\theta-\frac{\pi}{2}\right)^{2p}}}{\tilde{\rho}_{\mathrm{o}}^{2n+1}},\label{b1}\end{equation}
where $\Delta\equiv r-r_{\mathrm{o}}$, and $b_{nkp(a)}$ is the coefficient
of each individual term that depends on $n$, $k$ and $p$ as well
as $a$, with a dimension $\mathcal{R}^{k-1}$ for $a=t,\, r$ and
$\mathcal{R}^{k}$ for $a=\theta,\,\phi$. We recall from Eqs.~(\ref{eq:33})
and (\ref{eq:34}) that the first of the steps to lead to $\tilde{\rho}_{\mathrm{o}}^{2}$
in the denominator is replacing $\phi-\phi_{\mathrm{o}}$ by $\left(\phi-\phi'\right)-\frac{{J\dot{{r}}}}{f(r_{\mathrm{o}}^{2}+J^{2})}\Delta$
to eliminate the coupling term $\Delta\left(\phi-\phi_{\mathrm{o}}\right)$.
This makes a sum of independent square forms of each of $\Delta$
and $\phi-\phi'$, which is a necessary step to induce the Legendre
polynomial expansions later. Thus, to be consistent with this modification
in the denominator, $\phi-\phi_{\mathrm{o}}$ in the numerator on
the right hand side of Eq.~(\ref{b1}) should be also replaced by
$\left(\phi-\phi'\right)-\frac{{J\dot{{r}}}}{f(r_{\mathrm{o}}^{2}+J^{2})}\Delta$.
Then, this will create a number of additional terms apart from $\left(\phi-\phi'\right)^{m}$
when we expand the quantity $\left[\left(\phi-\phi'\right)-\frac{{J\dot{{r}}}}{f(r_{\mathrm{o}}^{2}+J^{2})}\Delta\right]$
raised, say, to the $m$-th power, and the computation will be very
complicated.

By analyzing the structure of the quantity on the right hand side
of Eq.~(\ref{b1}) one can prove that $\phi-\phi_{\mathrm{o}}$ may
be replaced just by $\phi-\phi'$ in the numerator without the term
$-\frac{{J\dot{{r}}}}{f(r_{\mathrm{o}}^{2}+J^{2})}\Delta$ (the same
idea is found in Mino, Nakano, and Sasaki \citet{mino-nakano-sasaki(02)}).
The verification follows. The behavior of the quantity on the right
hand side of Eq.~(\ref{b1}), according to the powers of each factor,
is\begin{equation}
Q_{a}[\epsilon^{-1}]\sim\tilde{\rho}_{\mathrm{o}}^{-(2n+1)}\Delta^{2n-k}\left(\phi-\phi_{\mathrm{o}}\right)^{k-2p}\left(\theta-\frac{\pi}{2}\right)^{2p}\mathrm{\mathcal{R}}^{s},\label{b2}\end{equation}
where $s=k-1$ for $a=t,\, r$ and $s=k$ for $a=\theta,\,\phi$.
Further,\begin{eqnarray}
\left(\phi-\phi_{\mathrm{o}}\right)^{k-2p} & = & \left[\left(\phi-\phi'\right)-\frac{{J\dot{{r}}\Delta}}{f(r_{\mathrm{o}}^{2}+J^{2})}\right]^{k-2p}\nonumber \\
 & = & \sum_{i=0}^{k-2p}c_{kpi}\left(\phi-\phi'\right)^{i}\Delta^{k-2p-i}\sim\left(\phi-\phi'\right)^{i}\Delta^{k-2p-i}/\mathcal{R}^{k-2p-i}\label{b3}\\
 & \sim & \left(\sin\Theta\right)^{i}\left(\cos\Phi\right)^{i}\Delta^{k-2p-i}/\mathcal{R}^{k-2p-i}+O[(x-x_{\mathrm{o}})^{k-2p+2}],\label{b4}\end{eqnarray}
where a binomial expansion over the index $i=0,\,1,\,\cdots\,,\, k-2p$
is assumed with $c_{kpi}\sim1/\mathcal{R}^{k-2p-i}$ in Eq.~(\ref{b3}),
and in Eq.~(\ref{b4}) $\left(\phi-\phi'\right)^{i}$ is replaced
by $[\sin(\phi-\phi')]^{i}+O[\left(\phi-\phi'\right)^{i+2}]$ ---
the term $O[(x-x_{\mathrm{o}})^{k-2p+2}]$ at the end results from
this $O[\left(\phi-\phi'\right)^{i+2}]$, then the coordinates are
rotated using the definition of new angles by Eq.~(\ref{eq:35}).
Also, by Eq.~(\ref{eq:35}) again\begin{equation}
\left(\theta-\frac{\pi}{2}\right)^{2p}=\left(\sin\Theta\right)^{2p}\left(\sin\Phi\right)^{2p}+O[(x-x_{\mathrm{o}})^{2p+2}].\label{b5}\end{equation}
Using Eqs.~(\ref{b4}) and (\ref{b5}), the behavior of $Q[\epsilon^{-1}]$
in Eq.~(\ref{b2}) looks like\begin{equation}
Q_{a}[\epsilon^{-1}]\sim\tilde{\rho}_{\mathrm{o}}^{-(2n+1)}\Delta^{2n-2p-i}\left(\sin\Theta\right)^{2p+i}\left(\cos\Phi\right)^{i}\left(\sin\Phi\right)^{2p}\mathcal{R}^{s},\label{b6}\end{equation}
where $s=2p+i-1$ for $a=t,\, r$ and $s=2p+i$ for $a=\theta,\,\phi$,
and any contributions from $O[(x-x_{\mathrm{o}})^{k-2p+2}]$ in Eq.~(\ref{b4})
and from $O[(x-x_{\mathrm{o}})^{2p+2}]$ in Eq.~(\ref{b5}) have
been disregarded: by putting these pieces into Eq.~(\ref{b2}) we
simply obtain $\epsilon^{1}$-terms, which would correspond to $O(\ell^{-2})$
in Eq.~(\ref{eq:9}) and should vanish when summed over $\ell$ in
our final self-force calculation by Eq.~(\ref{eq:8}) %
\footnote{Developing our mode-sum regularization scheme further, we may include
$\epsilon^{1}$-term in Eq.~(\ref{eq:30}) and this would generate
the next-order regularization terms $-2\sqrt{2}D_{a}/[(2\ell-1)(2\ell+3)]$
in the place of $O(\ell^{-2})$ in Eq.~(\ref{eq:9}). Rigorously,
their contributions to the self-force would be non-vanishing since
we are taking the sum of $-2\sqrt{2}D_{a}/[(2\ell-1)(2\ell+3)]$ over
many but finite number of $\ell$'s. Thorough discussions on $D_{a}$-terms
for a circular and for a general orbit cases are found in Ref.~\citet{detweiler-m-w(03)}
and Ref.~\citet{kim-detweiler(04)}, respectively.%
}. $Q_{a}[\epsilon^{-1}]$ then can be categorized into the following
cases:

\begin{enumerate}
\item $i=2j+1$ ($j=0,\,1,\,2,\,\cdots$) \\
The integrand for {}``$\left\langle \,\right\rangle $'' process,
$F(\Phi)\equiv\left(\cos\Phi\right)^{2j+1}\left(\sin\Phi\right)^{2p}$
has the property $F(\Phi+\pi)=-F(\Phi)$. Thus \begin{equation}
\left\langle Q_{a}[\epsilon^{-1}]\right\rangle =0,\label{b7}\end{equation}

\item $i=2j$ ($j=0,\,1,\,2,\,\cdots$)\\
Using Eqs.~(\ref{eq:37}) and (\ref{eq:39}), we can express $\left(\sin\Theta\right)^{2p+i}$
in Eq.~(\ref{b6}) above in terms of $\tilde{\rho}_{\mathrm{o}}$
and $\Delta$ via a binomial expansion \begin{eqnarray}
\left(\sin\Theta\right)^{2p+2j} & = & \left[2\left(1-\cos\Theta\right)\right]^{p+j}+O[(x-x_{\mathrm{o}})^{2(p+j)+2}]\nonumber \\
 & = & \sum_{q=0}^{p+j}d_{pjq}\tilde{\rho}_{\mathrm{o}}^{2q}\Delta^{2(p+j-q)}+O[(x-x_{\mathrm{o}})^{2(p+j)+2}]\\
 & \sim & \tilde{\rho}_{\mathrm{o}}^{2q}\Delta^{2(p+j-q)}/\mathcal{R}^{2(p+j)}+O[(x-x_{\mathrm{o}})^{2(p+j)+2}],\label{b13}\end{eqnarray}
 where $q=0,\,1,\,\cdots\,,\, p+j$ is the index for a binomial expansion
and $d_{pjq}\sim1/\mathcal{R}^{2(p+j)}$. When Eq.~(\ref{b13}) is
substituted into Eq.~(\ref{b6}), the contribution from $O[(x-x_{\mathrm{o}})^{2(p+j)+2}]$
can be disregarded since it would correspond to $O(\epsilon^{1})$
again. Then, we have \begin{equation}
Q_{a}[\epsilon^{-1}]\sim\left(\sin\Phi\right)^{2p}\left(\cos\Phi\right)^{2j}\tilde{\rho}_{\mathrm{o}}^{-2(n-q)-1}\Delta^{2(n-q)}\mathcal{R}^{s},\label{b8}\end{equation}
 where $s=-1$ for $a=t,\, r$ and $s=0$ for $a=\theta,\,\phi$,
and we can guarantee that $n-q\geq0$ always since $0\leq q\leq p+j=p+\frac{{1}}{2}i$,
$0\leq i\leq k-2p$, and $p\leq k\leq2n$. Then, Eq.~(\ref{b8})
can be subcategorized into the following two cases;

\begin{enumerate}
\item $n-q\geq1$ \\
By Eqs.~(\ref{eq:37}), (\ref{eq:39}), and (\ref{eq:42}) \begin{equation}
Q_{a}[\epsilon^{-1}]\begin{array}{c}
\\\sim\\
^{\Delta\rightarrow0}\end{array}\left(\sin\Phi\right)^{2p}\left(\cos\Phi\right)^{2j}\Delta P_{\ell}(\cos\Theta)\mathcal{R}^{s}\longrightarrow0,\label{b9}\end{equation}

\item $n-q=0$\\
By Eqs.~(\ref{eq:37}), (\ref{eq:39}), and (\ref{rho^-1/2}) \begin{equation}
Q_{a}[\epsilon^{-1}]\begin{array}{c}
\\\sim\\
^{\Delta\rightarrow0}\end{array}\left(\sin\Phi\right)^{2p}\left(\cos\Phi\right)^{2j}P_{\ell}(\cos\Theta)\mathcal{R}^{s},\label{b10}\end{equation}
where $s=-1$ for $a=t,\, r$ and $s=0$ for $a=\theta,\,\phi$.
\end{enumerate}
\end{enumerate}
Therefore, by analyzing the structure of $Q_{a}[\epsilon^{-1}]$ we
find that the $\epsilon^{-1}$-terms vanish in all the cases except
when $n-q=0$. The non-vanishing \emph{}$B_{a}$-terms are derived
only from this case. Then, by $0\leq q\leq p+j=p+\frac{{1}}{2}i$,
$0\leq i\leq k-2p$, and $p\leq k\leq2n$ together with $n=q$ one
can show that\begin{eqnarray}
0 & \leq & k-2p-i\;\mathrm{and}\; k-2p-i\leq0,\;\mathrm{i.e}.\; k-2p-i=0.\label{b11}\end{eqnarray}
Substituting this result into Eq.~(\ref{b3}), then into Eq.~(\ref{b1})
we may conclude that in the numerator of $Q[\epsilon^{-1}]$ in Eq.~(\ref{b1})
one can simply substitute \begin{equation}
\left(\phi-\phi_{\mathrm{o}}\right)^{k-2p}\rightarrow\left(\phi-\phi'\right)^{k-2p}.\;\mathrm{Q}.\,\mathrm{E.}\,\mathrm{D}.\label{b12}\end{equation}

The significance of this proof does not lie in the result given by
Eq.~(\ref{b12}) only, but also in the fact that the non-vanishing
contribution comes only from the case $n=q$ for Eq.~(\ref{b8}),
i.e.\begin{equation}
Q_{a}[\epsilon^{-1}]\sim\left(\sin\Phi\right)^{2p}\left(\cos\Phi\right)^{2(n-p)}\tilde{\rho}_{\mathrm{o}}^{-1}\mathcal{R}^{s},\label{b14}\end{equation}
where $n=1,\,2$ and $0\leq p\leq n$, and $s=-1$ for $a=t,\, r$
and $s=0$ for $a=\theta,\,\phi$.

Below are presented the calculations of $B_{a}$-terms of the regularization
parameters by component, in a similar manner to those for $A_{a}$-terms
.

\subsubsection{$B_{t}$-term\emph{:}}

We begin with\begin{equation}
Q_{t}[\epsilon^{-1}]=q^{2}\left\{ -\frac{{1}}{2}\frac{{\left.\partial_{t}\mathcal{P}_{\mathrm{III}}\right|_{t=t_{\mathrm{o}}}}}{\tilde{{\rho}}_{\mathrm{o}}^{3}}+\frac{{3}}{4}\frac{{\left.\left[\partial_{t}\left(\tilde{{\rho}}^{2}\right)\right]\mathcal{P}_{\mathrm{III}}\right|_{t=t_{\mathrm{o}}}}}{\tilde{{\rho}}_{\mathrm{o}}^{5}}\right\} .\label{eq:53}\end{equation}
The subsequent computation will be very lengthy and it will be reasonable
to split $Q_{t}[\epsilon^{-1}]$ into two parts. First, let \begin{equation}
Q_{t(1)}[\epsilon^{-1}]\equiv-\frac{{q^{2}}}{2}\tilde{{\rho}}_{\mathrm{o}}^{-3}\left.\partial_{t}\mathcal{P}_{\mathrm{III}}\right|_{t=t_{\mathrm{o}}},\label{eq:56}\end{equation}
where\begin{equation}
\left.\partial_{t}\mathcal{P}_{\mathrm{III}}\right|_{t=t_{\mathrm{o}}}=-\frac{{ME\dot{{r}}\Delta^{2}}}{f^{2}r_{\mathrm{o}}^{2}}-2\left(1-\frac{{M}}{r_{\mathrm{o}}}\right)\frac{{EJ\Delta}}{fr_{\mathrm{o}}}\left(\phi-\phi_{\mathrm{o}}\right)+r_{\mathrm{o}}E\dot{{r}}\left[\left(\phi-\phi_{\mathrm{o}}\right)^{2}+\left(\theta-\frac{{\pi}}{2}\right)^{2}\right].\label{eq:54}\end{equation}
As proved at the beginning of this Subsection, every $\left(\phi-\phi_{\mathrm{o}}\right)^{m}$
in the numerators of the $\epsilon^{-1}$-term can be replaced by
$\left(\phi-\phi'\right)^{m}$ without affecting the rest of calculation.
Then, followed by the rotation of the coordinates via Eq.~(\ref{eq:35})
\begin{eqnarray}
Q_{t(1)}[\epsilon^{-1}] & = & -\frac{{q^{2}}}{2}\tilde{{\rho}}_{\mathrm{o}}^{-3}\left[-\frac{{ME\dot{{r}}\Delta^{2}}}{f^{2}r_{\mathrm{o}}^{2}}-2\left(1-\frac{{M}}{r_{\mathrm{o}}}\right)\frac{{EJ\Delta}}{fr_{\mathrm{o}}}\sin\Theta\cos\Phi+2r_{\mathrm{o}}E\dot{{r}}\left(1-\cos\Theta\right)\right]\nonumber \\
 &  & +O\left[\frac{{(x-x_{\mathrm{o}})^{4}}}{\tilde{{\rho}}_{\mathrm{o}}^{3}}\right],\label{eq:57}\end{eqnarray}
where an approximation $\sin^{2}\Theta=2(1-\cos\Theta)+O[(x-x_{\mathrm{o}})^{4}]$
is used to obtain the last term inside the first bracket. Here we
may drop off the term $O\left[(x-x_{\mathrm{o}})^{4}/\tilde{{\rho}}_{\mathrm{o}}^{3}\right]$
, which is essentially $O(\epsilon^{1})$, for the same reason as
explained at the beginning of this subsection. Then, using the same
techniques as used to find $A_{a}$-terms, we can reduce Eq.~(\ref{eq:57})
to\begin{eqnarray}
Q_{t(1)}[\epsilon^{-1}] & = & \left[\frac{{q^{2}ME\dot{{r}}}}{2f^{2}r_{\mathrm{o}}^{2}}+\frac{{q^{2}r_{\mathrm{o}}^{3}E^{3}\dot{{r}}\chi^{-1}}}{2f^{2}\left(r_{\mathrm{o}}^{2}+J^{2}\right)^{2}}\right]\Delta^{2}\left[2\left(r_{\mathrm{o}}^{2}+J^{2}\right)\chi\left(\delta^{2}+1-\cos\Theta\right)\right]^{-3/2}\nonumber \\
 &  & -\frac{{q^{2}\left(1-\frac{{M}}{r_{\mathrm{o}}}\right)EJ\Delta\chi^{-3/2}\cos\Phi}}{\sqrt{2}fr_{\mathrm{o}}\left(r_{\mathrm{o}}^{2}+J^{2}\right)^{3/2}}\left.\frac{{\partial}}{\partial\Theta}\right|_{\Delta}\left(\delta^{2}+1-\cos\Theta\right)^{-1/2}-\frac{{q^{2}E\dot{{r}}r_{\mathrm{o}}\chi^{-1}}}{2\left(r_{\mathrm{o}}^{2}+J^{2}\right)}\tilde{{\rho}}_{\mathrm{o}}^{-1}.\label{eq:58}\end{eqnarray}
 As we have seen before, by Eq.~(\ref{eq:42}) $\left(\delta^{2}+1-\cos\Theta\right)^{-3/2}\sim\Delta^{-1}$
in the limit $\Delta\rightarrow0$ and the first term on the right
hand side will vanish. The second term will also give no contribution
to the regularization parameters because $\left\langle \chi^{-3/2}\cos\Phi\right\rangle =0$.
Only the last term, which is $\sim\tilde{{\rho}}_{\mathrm{o}}^{-1}$,
will give non-zero contribution according to the argument in the analysis
presented above (see Eq.~(\ref{b14})). Using Eq.~(\ref{rho^-1/2})
in the limit $\Delta\rightarrow0$ and taking {}``$\left\langle \,\right\rangle $''
process, Eq.~(\ref{eq:58}) becomes\begin{equation}
\left\langle \lim_{\Delta\rightarrow0}Q_{t(1)}[\epsilon^{-1}]\right\rangle =-\frac{{1}}{2}\frac{{q^{2}}}{r_{\mathrm{o}}^{2}}\frac{{E\dot{{r}}\left\langle \chi^{-3/2}\right\rangle }}{\left(1+J^{2}/r_{\mathrm{o}}^{2}\right)^{3/2}}\sum_{\ell=0}^{\infty}P_{\ell}\left(\cos\Theta\right).\label{eq:59}\end{equation}
 The identity $\left\langle \chi^{-p}\right\rangle \equiv\left\langle \left(1-\alpha\sin^{2}\Phi\right)^{-p}\right\rangle ={}_{2}F_{1}\left(p,\,\frac{{1}}{2};\,1,\,\alpha\right)\equiv F_{p}$,
with $\alpha\equiv J^{2}/\left(r_{\mathrm{o}}^{2}+J^{2}\right)$ is
taken from Appendix C of Paper I \citet{detweiler-m-w(03)}, and we
take the limit $\Theta\rightarrow0$ \begin{equation}
\left.\left\langle \lim_{\Delta\rightarrow0}Q_{t(1)}[\epsilon^{-1}]\right\rangle \right|_{\Theta\rightarrow0}=-\frac{{1}}{2}\frac{{q^{2}}}{r_{\mathrm{o}}^{2}}\frac{{E\dot{{r}}F_{3/2}}}{\left(1+J^{2}/r_{\mathrm{o}}^{2}\right)^{3/2}}.\label{eq:61}\end{equation}

Now the remaining part is\begin{equation}
Q_{t(2)}[\epsilon^{-1}]\equiv\frac{{3q^{2}}}{4}\tilde{{\rho}}_{\mathrm{o}}^{-5}\left.\left[\partial_{t}\left(\tilde{{\rho}}^{2}\right)\right]\mathcal{P}_{\mathrm{III}}\right|_{t=t_{\mathrm{o}}},\label{eq:62}\end{equation}
where

\begin{eqnarray}
\left.\left[\partial_{t}\left(\tilde{{\rho}}^{2}\right)\right]\mathcal{P}_{\mathrm{III}}\right|_{t=t_{\mathrm{o}}} & = & \left[-\frac{{2E\dot{{r}}\Delta}}{f}-2EJ\left(\phi-\phi_{\mathrm{o}}\right)\right]\nonumber \\
 &  & \times\left[-\left(1+\frac{{\dot{{r}}^{2}}}{f}\right)\frac{{M\Delta^{3}}}{f^{2}r_{\mathrm{o}}^{2}}+\left(2-\frac{{5M}}{r_{\mathrm{o}}}\right)\frac{{J\dot{{r}}\Delta^{2}}}{f^{2}r_{\mathrm{o}}}\left(\phi-\phi_{\mathrm{o}}\right)\right.\nonumber \\
 &  & +\left(1-\frac{{\dot{{r}}}}{f}+\frac{{2J^{2}}}{r_{\mathrm{o}}^{2}}\right)r_{\mathrm{o}}\Delta\left(\phi-\phi_{\mathrm{o}}\right)^{2}+\left(1-\frac{{\dot{{r}}^{2}}}{f}\right)r_{\mathrm{o}}\Delta\left(\theta-\frac{{\pi}}{2}\right)^{2}\nonumber \\
 &  & \left.-r_{\mathrm{o}}J\dot{{r}}\left(\phi-\phi_{\mathrm{o}}\right)^{3}-r_{\mathrm{o}}J\dot{{r}}\left(\phi-\phi_{\mathrm{o}}\right)\left(\theta-\frac{{\pi}}{2}\right)^{2}\right]+O[(x-x_{\mathrm{o}})^{6}].\label{eq:55}\end{eqnarray}
Taking similar procedures as above, the non-vanishing contributions
turn out to be\begin{eqnarray}
\left\langle \lim_{\Delta\rightarrow0}Q_{t(2)}[\epsilon^{-1}]\right\rangle  & = & \left\langle \lim_{\Delta\rightarrow0}\frac{{3}}{2}q^{2}EJ^{2}\dot{{r}}r_{\mathrm{o}}\tilde{{\rho}}_{\mathrm{o}}^{-5}\cos^{2}\Phi\sin^{4}\Theta\right\rangle \nonumber \\
 & = & \left\langle \lim_{\Delta\rightarrow0}\frac{{3}}{2}\frac{{q^{2}}}{r_{\mathrm{o}}}\frac{{E\dot{{r}}\tilde{{\rho}}_{\mathrm{o}}^{-1}}}{1+J^{2}/r_{\mathrm{o}}^{2}}\left(\chi^{-1}-\frac{{\chi^{-2}}}{1+J^{2}/r_{\mathrm{o}}^{2}}\right)\right\rangle \nonumber \\
 & = & \frac{{3}}{2}\frac{{q^{2}}}{r_{\mathrm{o}}^{2}}\frac{{E\dot{{r}}}}{\left(1+J^{2}/r_{\mathrm{o}}^{2}\right)^{3/2}}\left(\left\langle \chi^{-3/2}\right\rangle -\frac{{\left\langle \chi^{-5/2}\right\rangle }}{1+J^{2}/r_{\mathrm{o}}^{2}}\right)\sum_{\ell=0}^{\infty}P_{\ell}\left(\cos\Theta\right),\label{eq:63}\end{eqnarray}
where all other terms than $\sim\tilde{{\rho}}_{\mathrm{o}}^{-1}$
again have been dropped off during the procedure since they vanish
either in the limit $\Delta\rightarrow0$ or through the {}``$\left\langle \,\right\rangle $''
process. Then, using the identity $\left\langle \chi^{-p}\right\rangle \equiv{}_{2}F_{1}\left(p,\,\frac{{1}}{2};\,1,\,\alpha\right)\equiv F_{p}$,
we have\begin{equation}
\left.\left\langle \lim_{\Delta\rightarrow0}Q_{t(2)}[\epsilon^{-1}]\right\rangle \right|_{\Theta\rightarrow0}=\frac{{3}}{2}\frac{{q^{2}}}{r_{\mathrm{o}}^{2}}\frac{{E\dot{{r}}}}{\left(1+J^{2}/r_{\mathrm{o}}^{2}\right)^{3/2}}\left(F_{3/2}-\frac{{F_{5/2}}}{1+J^{2}/r_{\mathrm{o}}^{2}}\right).\label{eq:65}\end{equation}

By combining Eqs.~(\ref{eq:61}) and (\ref{eq:65}), we finally obtain\begin{equation}
B_{t}=\frac{{q^{2}}}{r_{\mathrm{o}}^{2}}E\dot{r}\left[\frac{{F_{3/2}}}{\left(1+J^{2}/r_{\mathrm{o}}^{2}\right)^{3/2}}-\frac{{3F_{5/2}}}{2\left(1+J^{2}/r_{\mathrm{o}}^{2}\right)^{5/2}}\right].\label{eq:66}\end{equation}

\subsubsection{$B_{r}$-term\emph{:}}

From Eq.~(\ref{eq:52}) we start with\begin{equation}
Q_{r}[\epsilon^{-1}]=q^{2}\left\{ -\frac{{1}}{2}\frac{{\left.\partial_{r}\mathcal{P}_{\mathrm{III}}\right|_{t=t_{\mathrm{o}}}}}{\tilde{{\rho}}_{\mathrm{o}}^{3}}+\frac{{3}}{4}\frac{{\left.\left[\partial_{r}\left(\tilde{{\rho}}^{2}\right)\right]\mathcal{P}_{\mathrm{III}}\right|_{t=t_{\mathrm{o}}}}}{\tilde{{\rho}}_{\mathrm{o}}^{5}}\right\} .\label{eq:67}\end{equation}
Then, following the same steps as taken for \textbf{}the case of $B_{t}$-term
\textbf{}above, we obtain\begin{equation}
B_{r}=\frac{{q^{2}}}{r_{\mathrm{o}}^{2}}\left[-\frac{{F_{1/2}}}{\left(1+J^{2}/r_{\mathrm{o}}^{2}\right)^{1/2}}+\frac{{(1-2f^{-1}\dot{{r}}^{2}){F_{3/2}}}}{2\left(1+J^{2}/r_{\mathrm{o}}^{2}\right)^{3/2}}+\frac{{3}f^{-1}\dot{{r}}^{2}F_{5/2}}{2\left(1+J^{2}/r_{\mathrm{o}}^{2}\right)^{5/2}}\right].\label{eq:68}\end{equation}

\subsubsection{$B_{\phi}$-term\emph{:}}

Again, from Eq.~(\ref{eq:52})\begin{equation}
Q_{\phi}[\epsilon^{-1}]=q^{2}\left\{ -\frac{{1}}{2}\frac{{\left.\partial_{\phi}\mathcal{P}_{\mathrm{III}}\right|_{t=t_{\mathrm{o}}}}}{\tilde{{\rho}}_{\mathrm{o}}^{3}}+\frac{{3}}{4}\frac{{\left.\left[\partial_{\phi}\left(\tilde{{\rho}}^{2}\right)\right]\mathcal{P}_{\mathrm{III}}\right|_{t=t_{\mathrm{o}}}}}{\tilde{{\rho}}_{\mathrm{o}}^{5}}\right\} .\label{eq:69}\end{equation}
Then, similarly we can derive\begin{equation}
B_{\phi}=\frac{{q^{2}}}{J}\dot{r}\left[\frac{{F_{1/2}-F_{3/2}}}{\left(1+J^{2}/r_{\mathrm{o}}^{2}\right)^{1/2}}+\frac{{3(F_{5/2}-F_{3/2})}}{2\left(1+J^{2}/r_{\mathrm{o}}^{2}\right)^{3/2}}\right].\label{eq:70}\end{equation}

\subsubsection{$B_{\theta}$-term\emph{:}}

As $A_{\theta}$ vanishes, so should $B_{\theta}$. From \begin{equation}
Q_{\theta}[\epsilon^{-1}]=q^{2}\left\{ -\frac{{1}}{2}\frac{{\left.\partial_{\theta}\mathcal{P}_{\mathrm{III}}\right|_{t=t_{\mathrm{o}}}}}{\tilde{{\rho}}_{\mathrm{o}}^{3}}+\frac{{3}}{4}\frac{{\left.\left[\partial_{\theta}\left(\tilde{{\rho}}^{2}\right)\right]\mathcal{P}_{\mathrm{III}}\right|_{t=t_{\mathrm{o}}}}}{\tilde{{\rho}}_{\mathrm{o}}^{5}}\right\} ,\label{eq:71}\end{equation}
one finds that there is no term like $\sim\tilde{{\rho}}_{\mathrm{o}}^{-1}$:
all terms are either like $\sim\Delta^{2n}/\tilde{{\rho}}_{\mathrm{o}}^{2n+1}$
or like $\sim\Delta^{2n-1}\sin\Theta\cos\Phi/\tilde{{\rho}}_{\mathrm{o}}^{2n+1}$
($n=1,\,2$), which vanish in the limit $\Delta\rightarrow0$ or through
the {}``$\left\langle \,\right\rangle $'' process. Thus\begin{equation}
B_{\theta}=0.\label{eq:72}\end{equation}

\subsection{$C_{a}$-terms\label{sub:C-terms}}

We have mentioned before that $C_{a}$-terms, which originate from
$\epsilon^{0}$-term in Eq.~(\ref{eq:30}), always vanish. This can
be proved by analyzing the structure of $\epsilon^{0}$-term. First
we specify the $\epsilon^{0}$-order term for$\left.\partial_{a}(1/\rho)\right|_{t=t_{\mathrm{o}}}$
in a Laurent series expansion and define\begin{equation}
Q_{a}[\epsilon^{0}]\equiv q^{2}\left\{ -\frac{{1}}{2}\frac{{\left.\partial_{a}\mathcal{P}_{\mathrm{IV}}\right|_{t=t_{\mathrm{o}}}}}{\tilde{{\rho}}_{\mathrm{o}}^{3}}+\frac{{3}}{4}\frac{{\left.\left(\partial_{a}\mathcal{P}_{\mathrm{III}}\right)\mathcal{P}_{\mathrm{III}}\right|_{t=t_{\mathrm{o}}}+\left.\left[\partial_{a}\left(\tilde{{\rho}}^{2}\right)\right]\mathcal{P}_{\mathrm{IV}}\right|_{t=t_{\mathrm{o}}}}}{\tilde{{\rho}}_{\mathrm{o}}^{5}}-\frac{{15}}{16}\frac{{\left.\left[\partial_{a}\left(\tilde{{\rho}}^{2}\right)\right]\mathcal{P}_{\mathrm{III}}^{2}\right|_{t=t_{\mathrm{o}}}}}{\tilde{{\rho}}_{\mathrm{o}}^{7}}\right\} .\label{eq:c0}\end{equation}
Generically, this can be written as

\begin{equation}
Q_{a}[\epsilon^{0}]=\sum_{n=1}^{3}\sum_{k=0}^{2n+1}\sum_{p=0}^{[k/2]}\frac{{c_{nkp(a)}\Delta^{2n+1-k}\left(\phi-\phi_{\mathrm{o}}\right)^{k-2p}\left(\theta-\frac{\pi}{2}\right)^{2p}}}{\tilde{\rho}_{\mathrm{o}}^{2n+1}},\label{c1}\end{equation}
 where $\Delta\equiv r-r_{\mathrm{o}}$, and $c_{nkp(a)}$ is the
coefficient of each individual term that depends on $n$, $k$ and
$p$ as well as $a$, with a dimension $\mathcal{R}^{k-2}$ for $a=t,\, r$
and $\mathcal{R}^{k-1}$ for $a=\theta,\,\phi$.

The behavior of $Q_{a}[\epsilon^{0}]$, according to the powers of
each factor on the right hand side of Eq.~(\ref{c1}), is\begin{equation}
Q_{a}[\epsilon^{0}]\sim\tilde{\rho}_{\mathrm{o}}^{-(2n+1)}\Delta^{2n+1-k}\left(\phi-\phi_{\mathrm{o}}\right)^{k-2p}\left(\theta-\frac{\pi}{2}\right)^{2p}\mathcal{R}^{s},\label{c2}\end{equation}
where $s=k-2$ for $a=t,\, r$ and $s=k-1$ for $a=\theta,\,\phi$.
Following the same procedure as in the beginning of Subsection~\ref{sub:B-terms},
Eq.~(\ref{c2}) becomes\begin{equation}
Q_{a}[\epsilon^{0}]\sim\tilde{\rho}_{\mathrm{o}}^{-(2n+1)}\Delta^{2n+1-2p-i}\left(\sin\Theta\right)^{2p+i}\left(\sin\Phi\right)^{2p}\left(\cos\Phi\right)^{i}\mathcal{R}^{s},\label{c3}\end{equation}
where a binomial expansion over the index $i=0,\,1,\,\cdots\,,\, k-2p$
is assumed, and $s=2p+i-2$ for $a=t,\, r$ and $s=2p+i-1$ for $a=\theta,\,\phi$.
Here we have disregarded any by-products like $O[(x-x_{\mathrm{o}})^{k-2p+2}]$
and $O[(x-x_{\mathrm{o}})^{2p+2}]$, which originate from $\left(\phi-\phi_{\mathrm{o}}\right)^{k-2p}$
and $\left(\theta-\frac{\pi}{2}\right)^{2p}$, respectively when we
rotate the angles: by putting them back into Eq.~(\ref{c2}) we simply
obtain $\epsilon^{2}$-terms, which would correspond to $O(\ell^{-4})$
in Eq.~(\ref{eq:9}) and should vanish when summed over $\ell$ in
our final self-force calculation by Eq.~(\ref{eq:8}). Then, the
rest of the argument is developed in the same way as in the beginning
of Subsection~\ref{sub:B-terms}:

\begin{enumerate}
\item $i=2j+1$ ($j=0,\,1,\,2,\,\cdots$) \\
The integrand for {}``$\left\langle \,\right\rangle $'' process,
$F(\Phi)\equiv\left(\cos\Phi\right)^{2j+1}\left(\sin\Phi\right)^{2p}$
has the property $F(\Phi+\pi)=-F(\Phi)$. Thus \begin{equation}
\left\langle Q_{a}[\epsilon^{0}]\right\rangle =0,\label{c4}\end{equation}

\item $i=2j$ ($j=0,\,1,\,2,\,\cdots$) \\
We have\begin{equation}
Q_{a}[\epsilon^{0}]\sim\left(\sin\Phi\right)^{2p}\left(\cos\Phi\right)^{2j}\tilde{\rho}_{\mathrm{o}}^{-2(n-q)-1}\Delta^{2(n-q)+1}\mathcal{R}^{s},\label{c5}\end{equation}
where $q=0,\,1,\,\cdots\,,\, p+j$ is the index for a binomial expansion
and $s=-2$ for $a=t,\, r$ and $s=-1$ for $a=\theta,\,\phi$. Here
we can guarantee that $n-q\geq-\frac{{1}}{2}$, i.e. $n-q=0,\,1,\,2,\,\cdots$
since $0\leq q\leq p+j=p+\frac{{1}}{2}i$, $0\leq i\leq k-2p$, and
$p\leq k\leq2n+1$. Then, Eq.~(\ref{c5}) can be subcategorized into
the following two cases;

\begin{enumerate}
\item $n-q\geq1$\\
By Eqs.~(\ref{eq:37}), (\ref{eq:39}), and (\ref{eq:42})\begin{equation}
Q_{a}[\epsilon^{0}]\begin{array}{c}
\\\sim\\
^{\Delta\rightarrow0}\end{array}\left(\sin\Phi\right)^{2p}\left(\cos\Phi\right)^{2j}\Delta^{2}P_{\ell}(\cos\Theta)\mathcal{R}^{s}\longrightarrow0,\label{c6}\end{equation}

\item $n-q=0$\\
By Eqs.~(\ref{eq:37}), (\ref{eq:39}), and (\ref{rho^-1/2}) \begin{equation}
Q_{a}[\epsilon^{0}]\begin{array}{c}
\\\sim\\
^{\Delta\rightarrow0}\end{array}\left(\sin\Phi\right)^{2p}\left(\cos\Phi\right)^{2j}\Delta P_{\ell}(\cos\Theta)\mathcal{R}^{s}\longrightarrow0,\label{c7}\end{equation}
where $s=-2$ for $a=t,\, r$ and $s=-1$ for $a=\theta,\,\phi$.
\end{enumerate}
\end{enumerate}
Clearly, in any cases the quantity $Q_{a}[\epsilon^{0}]$ does not
survive, therefore we can conclude that $C_{a}$-terms are always
zero.~Q. E. D. 

Also, this justifies the argument that we need not clarify the term
$O[(x-x_{\mathrm{o}})^{3}]$ in Eq.~(\ref{eq:20}) and its contribution
to $\rho^{2}$, which is $O[(x-x_{\mathrm{o}})^{4}]$ in Eqs.~(\ref{eq:26})
and (\ref{eq:27}) in Section \ref{sec:DETERMINATION-OF-psiS} or
$\mathcal{P}_{\mathrm{IV}}$ in Eqs.~(\ref{eq:29}) and (\ref{new_rho^2})
in Section \ref{sec:-REGULARIZATION-PARAMETERS}: by the analysis
of the generic structure given above, $-\frac{{1}}{2}\left.\partial_{a}\mathcal{P}_{\mathrm{IV}}\right|_{t=t_{\mathrm{o}}}/\tilde{{\rho}}_{\mathrm{o}}^{3}$
or $\frac{{3}}{4}\left.\left[\partial_{a}\left(\tilde{{\rho}}^{2}\right)\right]\mathcal{P}_{\mathrm{IV}}\right|_{t=t_{\mathrm{o}}}/\tilde{{\rho}}_{\mathrm{o}}^{5}$
would simply vanish in the coincidence limit $x\rightarrow x_{\mathrm{o}}$,
regardless of what $\mathcal{P}_{\mathrm{IV}}$ is.

\section{DISCUSSION}

\subsection*{The scalar field in the flat spacetime limit}

One interesting fact is that when we consider flat spacetime as the
background our singular source field $\psi^{\mathrm{S}}$ may be determined
in a completely different way from the case of curved spacetime. This
flat spacetime version of $\psi^{\mathrm{S}}$, though obtained using
a different method, should agree with its curved spacetime version
when the flat spacetime limit is taken. 

Without introducing the special coordinate frame like that of the
THZ coordinates as employed in Section~\ref{sec:DETERMINATION-OF-psiS},
this argument can be shown rather straightforward: the retarded field,
which is equivalent to the singular source field in flat spacetime,
can be computed directly using the coulomb potential and considering
the special relativity. Here we will focus on $\rho=\left|\vec{x}-\vec{x}_{\mathrm{o}}\right|$
between the field point $x$ and the source point $x_{\mathrm{o}}$.
This will be expressed in terms of the Cartesian coordinates first.
Then, we will switch from the Cartesian coordinates to the spherical
polar coordinates and arrange the terms in a polynomial according
to their orders, where quartic or higher order terms will be separated
as errors. Finally, the expression of $\rho^{2}$ for $\psi_{\mathrm{ret}}$
in flat spacetime obtained thus will be shown to agree with the flat
spacetime limit of $\rho^{2}$ for $\psi_{\mathrm{S}}$ in Eq.~(\ref{eq:27}). 

In the frame of reference in which the particle is always at rest
at the point $\vec{x_{\mathrm{o}}}$, the retarded field at $\vec{x}$
is \[
\psi_{\mathrm{ret}}=\frac{{q}}{\left|\vec{x}-\vec{x}_{\mathrm{o}}\right|}.\]
We boost this frame such that in the new frame of reference the particle
is moving with the 3-dim velocity $\vec{\beta}$ (see Ref.~\citet{jackson(99)}):

\begin{eqnarray*}
T\equiv t-t_{\mathrm{o}} & \;\Longrightarrow\; & T=\gamma\left(T'-\vec{\beta}\cdot\overrightarrow{X}'\right),\\
\overrightarrow{X}\equiv\vec{x}-\vec{x}_{\mathrm{o}} & \;\Longrightarrow\; & \overrightarrow{X}=\overrightarrow{X}'+\vec{\beta}\left[\frac{{\gamma-1}}{\beta^{2}}\left(\vec{\beta}\cdot\overrightarrow{X}'\right)-\gamma T'\right],\end{eqnarray*}
 where $\overrightarrow{X}'\equiv\vec{x}'-\vec{x}'_{\mathrm{o}}$
and $T'\equiv t'-t'_{\mathrm{o}}$. Then, we have \[
\rho^{2}\equiv\left|\vec{x}-\vec{x}_{\mathrm{o}}\right|^{2}=\left\{ \left(\vec{x}'-\vec{x}'_{\mathrm{o}}\right)+\vec{\beta}\left[\frac{{\gamma-1}}{\beta^{2}}\left[\vec{\beta}\cdot\left(\vec{x}'-\vec{x}'_{\mathrm{o}}\right)\right]-\gamma\left(t'-t_{\mathrm{o}}'\right)\right]\right\} ^{2}.\]
Dropping the $'$ notation and expanding the terms inside $\{\,\}$
out\[
\rho^{2}=\gamma^{2}\beta^{2}\left(t-t_{\mathrm{o}}\right)^{2}-2\gamma^{2}\left(t-t_{\mathrm{o}}\right)\left[\vec{\beta}\cdot\left(\vec{x}-\vec{x}_{\mathrm{o}}\right)\right]+\gamma^{2}\left[\vec{\beta}\cdot\left(\vec{x}-\vec{x}_{\mathrm{o}}\right)\right]^{2}+\left(\vec{x}-\vec{x}_{\mathrm{o}}\right)^{2}.\]

Now we need convert this expression into the spherical polar representation.
First, using the cosine law, the distance between $\vec{x}=(r,\,\theta,\,\phi)$
and $\vec{x}_{\mathrm{o}}=(r_{\mathrm{o}},\,\theta_{\mathrm{o}},\,\phi_{\mathrm{o}})$
can be expressed as\[
\left|\vec{x}-\vec{x}_{\mathrm{o}}\right|^{2}=r^{2}+r_{\mathrm{o}}^{2}-2rr_{\mathrm{o}}\cos\vartheta,\]
where\[
\cos\vartheta=\sin\theta\sin\theta_{\mathrm{o}}\cos\left(\phi-\phi_{\mathrm{o}}\right)+\cos\theta\cos\theta_{\mathrm{o}},\]
which is obvious from the trigonometric rule. In particular, for the
particle moving along an equatorial orbit ($\theta_{\mathrm{o}}=\pi/2$)
we may rewrite the above as\begin{eqnarray*}
\left|\vec{x}-\vec{x}_{\mathrm{o}}\right|^{2} & = & \left(r-r_{\mathrm{o}}\right)^{2}+r_{\mathrm{o}}\left(r-r_{\mathrm{o}}\right)\left(\theta-\frac{{\pi}}{2}\right)^{2}+r_{\mathrm{o}}\left(r-r_{\mathrm{o}}\right)\left(\phi-\phi_{\mathrm{o}}\right)^{2}+r_{\mathrm{o}}^{2}\left(\theta-\frac{{\pi}}{2}\right)^{2}\\
 &  & +r_{\mathrm{o}}^{2}\left(\phi-\phi_{\mathrm{o}}\right)^{2}+O[(\theta-\pi/2,\,\phi-\phi_{\mathrm{o}})^{4}],\end{eqnarray*}
where all the trigonometric functions of the small arguments are expanded
in Taylor series up to the cubic order.

To compute $\vec{\beta}\cdot\left(\vec{x}-\vec{x}_{\mathrm{o}}\right)$,
first of all one need change the basis vectors from $\{\mathbf{\hat{x}},\,\mathbf{\hat{y}},\,\mathbf{\hat{z}}\}$
to $\{\mathbf{\hat{r}_{\mathrm{o}}},\,\mathbf{\hat{\mathbf{\theta}}_{\mathrm{o}}},\,\mathbf{\hat{\mathbf{\phi}}_{\mathrm{o}}}\}$
via\[
\left(\begin{array}{c}
\mathbf{\hat{x}}\\
\mathbf{\hat{y}}\\
\mathbf{\hat{z}}\end{array}\right)=\left(\begin{array}{ccc}
\cos\phi_{\mathrm{o}} & 0 & -\sin\phi_{\mathrm{o}}\\
\sin\phi_{\mathrm{o}} & 0 & \cos\phi_{\mathrm{o}}\\
0 & -1 & 0\end{array}\right)\left(\begin{array}{c}
\mathbf{\hat{r}_{\mathrm{o}}}\\
\mathbf{\mathbf{\hat{\mathbf{\theta}}}_{\mathrm{o}}}\\
\mathbf{\mathbf{\hat{\mathbf{\phi}}}_{\mathrm{o}}}\end{array}\right).\]
Then, $\vec{x}-\vec{x}_{\mathrm{o}}$ can be rewritten as\begin{eqnarray*}
\vec{x}-\vec{x}_{\mathrm{o}} & = & \left(r\sin\theta\cos\phi\right)\left(\mathbf{\hat{r}_{\mathrm{o}}}\cos\phi_{\mathrm{o}}-\mathbf{\hat{\phi}_{\mathrm{o}}\sin\phi_{\mathrm{o}}}\right)+\left(r\sin\theta\sin\phi\right)\left(\mathbf{\hat{r}_{\mathrm{o}}}\sin\phi_{\mathrm{o}}+\mathbf{\hat{\phi}_{\mathrm{o}}\cos\phi_{\mathrm{o}}}\right)\\
 &  & -r\cos\theta\hat{\mathbf{\theta}}_{\mathrm{o}}-r_{\mathrm{o}}\mathbf{\hat{r}_{\mathrm{o}}}.\end{eqnarray*}
Also, in the new basis the 3-dim velocity of the particle moving in
the equatorial plane is expressed as\[
\vec{\beta}=\beta_{r}\mathbf{\hat{r}_{\mathrm{o}}}+\beta_{\phi}\mathbf{\mathbf{\hat{\mathbf{\phi}}}_{\mathrm{o}}}.\]
Thus, taking a dot product of $\vec{\beta}$ and $\vec{x}-\vec{x}_{\mathrm{o}}$
gives\[
\vec{\beta}\cdot\left(\vec{x}-\vec{x}_{\mathrm{o}}\right)=\beta_{r}\left[r\sin\theta\cos\left(\phi-\phi_{\mathrm{o}}\right)-r_{\mathrm{o}}\right]+\beta_{\phi}r\sin\theta\sin\left(\phi-\phi_{\mathrm{o}}\right).\]
Again, taking a series expansion of this quantity around the source
point $\vec{x}_{\mathrm{o}}=\left(r_{\mathrm{o}},\,\frac{{\pi}}{2},\,\phi_{\mathrm{o}}\right)$,
sufficiently up to the quadratic order, we have \begin{eqnarray*}
\vec{\beta}\cdot\left(\vec{x}-\vec{x}_{\mathrm{o}}\right) & = & \beta_{r}\left(r-r_{\mathrm{o}}\right)+\beta_{\phi}r_{\mathrm{o}}\left(\phi-\phi_{\mathrm{o}}\right)+\beta_{\phi}\left(r-r_{\mathrm{o}}\right)\left(\phi-\phi_{\mathrm{o}}\right)-\frac{{1}}{2}r_{\mathrm{o}}\beta_{r}\left(\theta-\frac{{\pi}}{2}\right)^{2}\\
 &  & -\frac{{1}}{2}r_{\mathrm{o}}\beta_{r}\left(\phi-\phi_{\mathrm{o}}\right)^{2}+O[(\theta-\pi/2,\,\phi-\phi_{\mathrm{o}})^{3}].\end{eqnarray*}

Finally, putting all the terms together we obtain the following\begin{eqnarray*}
\rho^{2} & = & \left(E^{2}-1\right)\left(t-t_{\mathrm{o}}\right)^{2}-2E\dot{{r}}\left(t-t_{\mathrm{o}}\right)\left(r-r_{\mathrm{o}}\right)-2EJ\left(t-t_{\mathrm{o}}\right)\left(\phi-\phi_{\mathrm{o}}\right)\\
 &  & +\left(1+\dot{{r}}^{2}\right)\left(r-r_{\mathrm{o}}\right)^{2}+2J\dot{{r}}\left(r-r_{\mathrm{o}}\right)\left(\phi-\phi_{\mathrm{o}}\right)+\left(r_{\mathrm{o}}^{2}+J^{2}\right)\left(\phi-\phi_{\mathrm{o}}\right)^{2}+r_{\mathrm{o}}^{2}\left(\theta-\frac{{\pi}}{2}\right)^{2}\\
 &  & -\frac{2EJ}{r_{\mathrm{o}}}\left(t-t_{\mathrm{o}}\right)\left(r-r_{\mathrm{o}}\right)\left(\phi-\phi_{\mathrm{o}}\right)+r_{\mathrm{o}}E\dot{{r}}\left(t-t_{\mathrm{o}}\right)\left(\phi-\phi_{\mathrm{o}}\right)^{2}+r_{\mathrm{o}}E\dot{{r}}\left(t-t_{\mathrm{o}}\right)\left(\theta-\frac{{\pi}}{2}\right)^{2}\\
 &  & +\frac{{2J\dot{{r}}}}{r_{\mathrm{o}}}\left(r-r_{\mathrm{o}}\right)^{2}\left(\phi-\phi_{\mathrm{o}}\right)+r_{\mathrm{o}}\left(1-\dot{{r}}^{2}+\frac{{2J^{2}}}{r_{\mathrm{o}}^{2}}\right)\left(r-r_{\mathrm{o}}\right)\left(\phi-\phi_{\mathrm{o}}\right)^{2}\\
 &  & +r_{\mathrm{o}}\left(1-\dot{{r}}^{2}\right)\left(r-r_{\mathrm{o}}\right)\left(\theta-\frac{{\pi}}{2}\right)^{2}-r_{\mathrm{o}}J\dot{{r}}\left(\phi-\phi_{\mathrm{o}}\right)^{3}-r_{\mathrm{o}}J\dot{{r}}\left(\phi-\phi_{\mathrm{o}}\right)\left(\theta-\frac{{\pi}}{2}\right)^{2}\\
 &  & +O[(t-t_{\mathrm{o}},\, r-r_{\mathrm{o}},\,\theta-\pi/2,\,\phi-\phi_{\mathrm{o}})^{4}].\end{eqnarray*}
 This is exactly equal to the flat spacetime limit of $\rho^{2}$
for $\psi^{\mathrm{\mathrm{S}}}$ when $M=0$ in Eq.~(\ref{eq:27}),
with appropriate replacement for the coefficients, using $E=\left(dt/d\tau\right)_{\mathrm{o}}=\gamma$,
$J=r_{\mathrm{o}}^{2}\left(d\phi/d\tau\right)_{\mathrm{o}}=\gamma r_{\mathrm{o}}\beta_{\phi}$,
and $\dot{r}=\left(dr/d\tau\right)_{\mathrm{o}}=\gamma\beta_{r}$
together with the identity $\dot{{r}}^{2}=E^{2}-f\left(1+J^{2}/r_{\mathrm{o}}^{2}\right)$.

\begin{acknowledgments}
I would like to thank Professor Steven Detweiler for many helpful
suggestions and stimulating discussions during the course of this
project and in preparing this manuscript. I would also like to thank
Professor Bernard Whiting and Professor Richard Woodard for many useful
discussions. 
\end{acknowledgments}
\appendix

\section{THE THZ NORMAL COORDINATES\label{app.A.THZ}}

In Section~\ref{sec:DETERMINATION-OF-psiS} we have introduced the
Thorne-Hartle-Zhang's normal coordinates to simplify the description
of $\psi^{\mathrm{S}}$ as if it was measured by an observer who travels
on a particle moving in curved spacetime: in this coordinate system
$\{\mathcal{X}^{A}\}$($A=0,\,1,\,2,\,3$), $\psi^{\mathrm{S}}=q/\rho+O(\rho^{2}/\mathcal{R}^{3})$,
where $\rho^{2}\equiv\delta_{IJ}\mathcal{X}^{I}\mathcal{X}^{J}$ ($I,\, J=1,\,2,\,3$).
As presented by Eq.~(\ref{eq:27}), the specification of $O[(x-x_{\mathrm{o}})^{4}]$
in $\rho^{2}$ (thus of $O[(x-x_{\mathrm{o}})^{3}]$ in $\mathcal{X}^{I}$
in Eq.~(\ref{eq:20})) is not necessary for our current mode-sum
regularization scheme: it proves not to contribute to the regularization
parameters $A_{a}$, $B_{a}$, and $C_{a}$. 

In the practical calculation of self-force, however, it will be very
useful to extend our scheme to the next orders, i.e. to $D_{a}$ or
higher terms. If we extend Eq.~(\ref{eq:30}), say, to $\epsilon^{1}$-term,
it would generate the next-order regularization terms $-2\sqrt{2}D_{a}/[(2\ell-1)(2\ell+3)]$
in the place of $O(\ell^{-2})$ in Eq.~(\ref{eq:9}). Strictly, these
terms would give non-vanishing contributions to the self-force since
the sum of $-2\sqrt{2}D_{a}/[(2\ell-1)(2\ell+3)]$ is taken over many
but finite number of $\ell$'s in actual numerical calculations. 

In order to determine $D_{a}$ or higher terms, the knowledge of $O[(x-x_{\mathrm{o}})^{5}]$
in $\rho^{2}$ and thus of $O[(x-x_{\mathrm{o}})^{4}]$ in $\mathcal{X}^{I}$
will be required. As it should be an important tool for this purpose,
we present below a more detailed description of the THZ normal coordinates
$\{\mathcal{X}^{A}\}$($A=0,\,1,\,2,\,3$) for a particle moving along
a general orbit (confined to the equatorial plane for convenience)
about a Schwarzschild black hole %
\footnote{For derivations of the THZ-coordinates and for discussions on their
application to $D_{a}$-terms, readers may be referred to Ref.~\citet{detweiler-m-w(03)}
and Ref.~\citet{kim-detweiler(04)} for a circular and for a general
orbit cases, respectively.%
}: the expressions are given in terms of the Schwarzschild coordinates
$x^{a}=(t,\, r,\,\theta,\,\phi)$ and specified up to the quortic
order\\
\begin{equation}
\left\{ \begin{array}{l}
\mathcal{T}\equiv\mathcal{X}^{0}=-u_{\mathrm{o}A}\left[X^{A}+\alpha^{A}{}_{BCD}X^{B}X^{C}X^{D}+\beta^{A}{}_{BCDE}X^{B}X^{C}X^{D}X^{E}\right]+O(X^{5}),\\
\mathcal{X}^{I}=n_{\mathrm{o}A}^{(I)}\left[X^{A}+\kappa^{A}{}_{BCD}X^{B}X^{C}X^{D}+\lambda^{A}{}_{BCDE}X^{B}X^{C}X^{D}X^{E}\right]+O(X^{5})\end{array}\right.\label{a71}\end{equation}
with\begin{equation}
X^{A}=M^{A}{}_{a}(x^{a}-x_{\mathrm{o}}^{a})+{\displaystyle \frac{1}{2}}M^{A}{}_{a}\left.\Gamma_{bc}^{a}\right|_{\mathrm{o}}(x^{b}-x_{\mathrm{o}}^{b})(x^{c}-x_{\mathrm{o}}^{c})+O[(x-x_{\mathrm{o}})^{3}],\label{X^A}\end{equation}

\begin{eqnarray}
\alpha^{A}{}_{BCD} & \equiv & \frac{{1}}{6}\left.\Gamma_{PQ,R}^{A}\right|_{\mathrm{o}}\left(\pi^{P}{}_{B}\pi^{Q}{}_{C}\pi^{R}{}_{D}+3\pi^{Q}{}_{B}\pi^{R}{}_{C}h^{P}{}_{D}\right.\nonumber \\
 &  & \left.+3\pi^{R}{}_{B}h^{P}{}_{C}h^{Q}{}_{D}+h^{P}{}_{B}h^{Q}{}_{C}h^{R}{}_{D}\right),\label{alpha}\end{eqnarray}
\begin{eqnarray}
\beta^{A}{}_{BCDE} & \equiv & \frac{{1}}{24}\left.\Gamma_{PQ,RS}^{A}\right|_{\mathrm{o}}\left(\pi^{P}{}_{B}\pi^{Q}{}_{C}\pi^{R}{}_{D}\pi^{S}{}_{E}+4\pi^{Q}{}_{B}\pi^{R}{}_{C}\pi^{S}{}_{D}h^{P}{}_{E}\right.\nonumber \\
 &  & \left.6\pi^{R}{}_{B}\pi^{S}{}_{C}h^{P}{}_{D}h^{Q}{}_{E}+4\pi^{S}{}_{B}h^{P}{}_{C}h^{Q}{}_{D}h^{R}{}_{E}+h^{P}{}_{B}h^{Q}{}_{C}h^{R}{}_{D}h^{S}{}_{E}\right)\nonumber \\
 &  & -\frac{5}{168}\left.R^{A}{}_{PQR,S}\right|_{\mathrm{o}}\pi^{QS}h^{P}{}_{B}h^{R}{}_{C}h_{DE},\label{beta}\end{eqnarray}
\begin{eqnarray}
\kappa^{A}{}_{BCD} & \equiv & \frac{{1}}{6}\left.\Gamma_{PQ,R}^{A}\right|_{\mathrm{o}}\left(\pi^{P}{}_{B}\pi^{Q}{}_{C}\pi^{R}{}_{D}+3\pi^{Q}{}_{B}\pi^{R}{}_{C}h^{P}{}_{D}\right.\nonumber \\
 &  & \left.+3\pi^{R}{}_{B}h^{P}{}_{C}h^{Q}{}_{D}+h^{P}{}_{B}h^{Q}{}_{C}h^{R}{}_{D}\right)\nonumber \\
 &  & -\left.R^{E}{}_{PQR}\right|_{\mathrm{o}}\left(\frac{1}{6}\delta^{A}{}_{E}\pi^{PQ}h^{R}{}_{B}h_{CD}+\frac{1}{3}\delta^{A}{}_{B}\pi^{Q}{}_{E}h^{P}{}_{C}h^{R}{}_{D}\right),\label{kappa}\end{eqnarray}
\begin{eqnarray}
\lambda^{A}{}_{BCDE} & \equiv & \frac{{1}}{24}\left.\Gamma_{PQ,RS}^{A}\right|_{\mathrm{o}}\left(\pi^{P}{}_{B}\pi^{Q}{}_{C}\pi^{R}{}_{D}\pi^{S}{}_{E}+4\pi^{Q}{}_{B}\pi^{R}{}_{C}\pi^{S}{}_{D}h^{P}{}_{E}\right.\nonumber \\
 &  & \left.6\pi^{R}{}_{B}\pi^{S}{}_{C}h^{P}{}_{D}h^{Q}{}_{E}+4\pi^{S}{}_{B}h^{P}{}_{C}h^{Q}{}_{D}h^{R}{}_{E}+h^{P}{}_{B}h^{Q}{}_{C}h^{R}{}_{D}h^{S}{}_{E}\right)\nonumber \\
 &  & -\left.R^{F}{}_{PQR,S}\right|_{\mathrm{o}}\left(\frac{1}{6}\delta^{A}{}_{F}\pi^{PQ}\pi^{S}{}_{B}h^{R}{}_{C}h_{DE}+\frac{1}{3}\delta^{A}{}_{B}\pi^{Q}{}_{F}\pi^{S}{}_{C}h^{P}{}_{D}h^{R}{}_{E}\right.\nonumber \\
 &  & \left.+\frac{1}{24}\delta^{A}{}_{F}\pi^{PQ}h^{R}{}_{B}h^{S}{}_{C}h_{DE}+\frac{1}{24}\delta^{A}{}_{B}\pi^{Q}{}_{F}h^{P}{}_{C}h^{R}{}_{D}h^{S}{}_{E}+\frac{2}{63}\delta^{A}{}_{F}\pi^{RS}h^{P}{}_{B}h^{Q}{}_{C}h_{DE}\right),\label{lambda}\end{eqnarray}
where $A,\,\cdots,\, E$ and $P,\,\cdots,\, S=0,\,1,\,2,\,3$, and
$I,\, J,\, K,\, L=1,\,2,\,3$. For Eq.~(\ref{a71}) we have\begin{eqnarray}
u_{\mathrm{o}}^{A} & = & \left(f^{-1/2}E,\, f^{-1/2}\dot{{r}},\,\frac{{J}}{r_{\mathrm{o}}},\,0\right),\label{a74}\\
n_{\mathrm{o}}^{(1)A} & = & \left(-f^{-1/2}\dot{{r}},\,1+\frac{{\dot{{r}}^{2}}}{f^{1/2}E+f},\,\frac{{J\dot{{r}}}}{r_{\mathrm{o}}\left(E+f^{1/2}\right)},\,0\right),\label{a75}\\
n_{\mathrm{o}}^{(2)A} & = & \left(-\frac{{J}}{r_{\mathrm{o}}},\,\frac{{J\dot{{r}}}}{r_{\mathrm{o}}\left(E+f^{1/2}\right)},\,1+\frac{{J^{2}}}{r_{\mathrm{o}}^{2}\left(f^{-1/2}E+1\right)},\,0\right),\label{a76}\\
n_{\mathrm{o}}^{(3)A} & = & \left(0,\,0,\,0,\,1\right),\label{a77}\end{eqnarray}
where $f=\left(1-\frac{{2M}}{r_{\mathrm{o}}}\right)$, and $E\equiv-u_{t}=\left(1-2M/r_{\mathrm{o}}\right)\left(dt/d\tau\right)_{\mathrm{o}}$
($\tau$: proper time) and $J\equiv u_{\phi}=r_{\mathrm{o}}^{2}\left(d\phi/d\tau\right)_{\mathrm{o}}$
are the conserved energy and angular momentum in the background, respectively,
and $\dot{r}\equiv u^{r}=\left(dr/d\tau\right)_{\mathrm{o}}$. In
Eq.~(\ref{X^A}) we define\begin{equation}
x_{\mathrm{o}}^{a}=\left(t_{\mathrm{o}},\, r_{\mathrm{o}},\,\frac{{\pi}}{2},\,\phi_{\mathrm{o}}\right),\label{a78}\end{equation}
\begin{equation}
M^{A}{}_{a}=\mathrm{diag}\left[f^{1/2},\, f^{-1/2},\, r_{\mathrm{o}},\,-r_{\mathrm{o}}\right],\label{a79}\end{equation}
along with the non-zero Christoffel symbols at $x_{\mathrm{o}}$ in
the Schwarzschild background 

\begin{equation}
\left.\Gamma_{tr}^{t}\right|_{\mathrm{o}}=\frac{{M}}{fr_{\mathrm{o}}^{2}},\,\left.\Gamma_{tt}^{r}\right|_{\mathrm{o}}=\frac{{fM}}{r_{\mathrm{o}}^{2}},\,\left.\Gamma_{rr}^{r}\right|_{\mathrm{o}}=-\frac{{M}}{fr_{\mathrm{o}}^{2}},\,\left.\Gamma_{\theta\theta}^{r}\right|_{\mathrm{o}}=-fr_{\mathrm{o}},\,\left.\Gamma_{\phi\phi}^{r}\right|_{\mathrm{o}}=-fr_{\mathrm{o}},\,\left.\Gamma_{r\theta}^{\theta}\right|_{\mathrm{o}}=\left.\Gamma_{r\phi}^{\phi}\right|_{\mathrm{o}}=\frac{{1}}{r_{\mathrm{o}}}.\label{christoffel}\end{equation}
The quantities $\left.\Gamma_{BC,D}^{A}\right|_{\mathrm{o}}$, $\left.R^{A}{}_{BCD}\right|_{\mathrm{o}}$,
$\left.\Gamma_{BC,DE}^{A}\right|_{\mathrm{o}}$ and $\left.R^{A}{}_{BCD,E}\right|_{\mathrm{o}}$
in Eqs.~(\ref{alpha})-(\ref{lambda}) are evaluated from the initial
static normal coordinates $\{ X^{A}\}$ represented by Eq.~(\ref{X^A}).
They follow the identities\begin{eqnarray}
\left.\Gamma_{BC,D}^{A}\right|_{\mathrm{o}} & = & H^{A}{}_{BCD}+H^{A}{}_{CBD}-H_{BC}{}^{A}{}_{D},\label{G_ABCD}\\
\left.R^{A}{}_{BCD}\right|_{\mathrm{o}} & = & H_{BC}{}^{A}{}_{D}-H^{A}{}_{CBD}-H_{BD}{}^{A}{}_{C}+H^{A}{}_{DBC},\label{R_ABCD}\\
\left.\Gamma_{BC,DE}^{A}\right|_{\mathrm{o}} & = & 3\left(H^{A}{}_{BCDE}+H^{A}{}_{CBDE}-H_{BC}{}^{A}{}_{DE}\right),\label{G_ABCDE}\\
\left.R^{A}{}_{BCD,E}\right|_{\mathrm{o}} & = & 3\left(H_{BC}{}^{A}{}_{DE}-H^{A}{}_{CBDE}-H_{BD}{}^{A}{}_{CE}+H^{A}{}_{DBCE}\right),\label{R_ABCDE}\end{eqnarray}
 where the building blocks $H_{ABCD}$ and $H_{ABCDE}$ are taken
from $g_{AB}=\eta_{AB}+H_{ABCD}X^{C}X^{D}+H_{ABCDE}X^{C}X^{D}X^{E}$
(the linearized gravity in the geometry of $\{ X^{A}\}$) and have
symmetric properties $H_{ABCD}=H_{(AB)(CD)}$ and $H_{ABCDE}=H_{(AB)(CDE)}$.
The non-zero $H_{ABCD}$ and $H_{ABCDE}$ turn out to be 

\[
H_{0000}=-\frac{{M^{2}}}{fr_{\mathrm{o}}^{4}},\, H_{0011}=\frac{{1}}{f}\left(\frac{{2M}}{r_{\mathrm{o}}^{3}}-\frac{{3M^{2}}}{r_{\mathrm{o}}^{4}}\right),\, H_{0101}=\frac{{M^{2}}}{fr_{\mathrm{o}}^{4}},\, H_{0202}=H_{0303}=\frac{{M}}{2r_{\mathrm{o}}^{3}},\]
\[
H_{1100}=\frac{{M^{2}}}{fr_{\mathrm{o}}^{4}},\, H_{1111}=\frac{{1}}{f}\left(\frac{{2M}}{r_{\mathrm{o}}^{3}}-\frac{{M^{2}}}{r_{\mathrm{o}}^{4}}\right),\, H_{1122}=H_{1133}=-\frac{{f}}{r_{\mathrm{o}}^{2}},\, H_{1212}=H_{1313}=-\frac{{1}}{2}\left(\frac{{1}}{r_{\mathrm{o}}^{2}}-\frac{{M}}{r_{\mathrm{o}}^{3}}\right),\]
\[
H_{2222}=-\frac{{f}}{r_{\mathrm{o}}^{2}},\, H_{2233}=-\frac{{1}}{r_{\mathrm{o}}^{2}},\, H_{2323}=-\frac{{f}}{2r_{\mathrm{o}}^{2}},\, H_{3333}=-\frac{{f}}{r_{\mathrm{o}}^{2}},\]
and\[
H_{00001}=-\frac{1}{3f^{3/2}}\left(\frac{2M^{2}}{r_{\mathrm{o}}^{5}}-\frac{3M^{3}}{r_{\mathrm{o}}^{6}}\right),\, H_{00111}=-\frac{1}{f^{3/2}}\left(\frac{2M}{r_{\mathrm{o}}^{4}}-\frac{6M^{2}}{r_{\mathrm{o}}^{5}}+\frac{5M^{3}}{r_{\mathrm{o}}^{6}}\right),\,\]
\[
H_{00122}=H_{00133}=\frac{1}{3f^{1/2}}\left(\frac{2M}{r_{\mathrm{o}}^{4}}-\frac{3M^{2}}{r_{\mathrm{o}}^{5}}\right),\, H_{01000}=-\frac{M^{3}}{f^{3/2}r_{\mathrm{o}}^{6}},\, H_{01011}=-\frac{1}{3f^{3/2}}\left(\frac{4M^{2}}{r_{\mathrm{o}}^{5}}-\frac{3M^{3}}{r_{\mathrm{o}}^{6}}\right),\]
\[
H_{01022}=H_{01033}=\frac{M^{2}}{3f^{1/2}r_{\mathrm{o}}^{5}},\, H_{02012}=H_{03013}=-\frac{1}{6f^{1/2}}\left(\frac{M}{r_{\mathrm{o}}^{4}}-\frac{3M^{2}}{r_{\mathrm{o}}^{5}}\right),\]
\[
H_{11001}=-\frac{1}{3f^{3/2}}\left(\frac{2M^{2}}{r_{\mathrm{o}}^{5}}+\frac{3M^{3}}{r_{\mathrm{o}}^{6}}\right),\, H_{11111}=-\frac{1}{f^{3/2}}\left(\frac{2M}{r_{\mathrm{o}}^{4}}-\frac{6M^{2}}{r_{\mathrm{o}}^{5}}+\frac{3M^{3}}{r_{\mathrm{o}}^{6}}\right),\]
\[
H_{11122}=H_{11133}=\frac{1}{3f^{1/2}}\left(\frac{4}{r_{\mathrm{o}}^{3}}-\frac{14M}{r_{\mathrm{o}}^{4}}+\frac{15M^{2}}{r_{\mathrm{o}}^{5}}\right),\, H_{12002}=\frac{1}{6f^{1/2}}\left(\frac{M}{r_{\mathrm{o}}^{4}}-\frac{M^{2}}{r_{\mathrm{o}}^{5}}\right),\]
\[
H_{12112}=\frac{1}{6f^{1/2}}\left(\frac{4}{r_{\mathrm{o}}^{3}}-\frac{11M}{r_{\mathrm{o}}^{4}}+\frac{9M^{2}}{r_{\mathrm{o}}^{5}}\right),\, H_{12222}=-\frac{f^{1/2}}{2}\left(\frac{1}{r_{\mathrm{o}}^{3}}-\frac{M}{r_{\mathrm{o}}^{4}}\right),\, H_{12233}=\frac{f^{1/2}}{6}\left(\frac{1}{r_{\mathrm{o}}^{3}}+\frac{M}{r_{\mathrm{o}}^{4}}\right),\]
\[
H_{13003}=\frac{1}{6f^{1/2}}\left(\frac{M}{r_{\mathrm{o}}^{4}}-\frac{M^{2}}{r_{\mathrm{o}}^{5}}\right),\, H_{13113}=\frac{1}{6f^{1/2}}\left(\frac{4}{r_{\mathrm{o}}^{3}}-\frac{11M}{r_{\mathrm{o}}^{4}}+\frac{9M^{2}}{r_{\mathrm{o}}^{5}}\right),\]
\[
H_{13223}=-\frac{f^{1/2}}{6}\left(\frac{1}{r_{\mathrm{o}}^{3}}-\frac{M}{r_{\mathrm{o}}^{4}}\right),\, H_{13333}=-\frac{f^{1/2}}{2}\left(\frac{1}{r_{\mathrm{o}}^{3}}-\frac{M}{r_{\mathrm{o}}^{4}}\right),\]
\[
H_{22122}=H_{33133}=\frac{2f^{1/2}}{3}\left(\frac{1}{r_{\mathrm{o}}^{3}}-\frac{3M}{r_{\mathrm{o}}^{4}}\right),\, H_{22133}=\frac{2f^{1/2}}{3r_{\mathrm{o}}^{3}},\, H_{23123}=\frac{f^{1/2}}{3}\left(\frac{1}{r_{\mathrm{o}}^{3}}-\frac{3M}{r_{\mathrm{o}}^{4}}\right).\]

According to Ref.~(\citet{weinberg(72)}), one can show the following
out of the two geometries, the THZ $\{\mathcal{X}^{A}\}$ and the
Schwarzschild $\{ x^{a}\}$\begin{eqnarray}
\tilde{g}^{AB} & = & g^{ab}\frac{{\partial\mathcal{X}^{A}}}{\partial x^{a}}\frac{{\partial\mathcal{X}^{B}}}{\partial x^{b}}\nonumber \\
 & = & \eta^{AB}+O[(x-x_{\mathrm{o}})^{2}]\nonumber \\
 & = & \eta^{AB}+O(\mathcal{X}^{2}),\label{eq:a80}\end{eqnarray}
where tilde denotes the THZ geometry and $O[(x-x_{\mathrm{o}})^{2}]$
is converted to $O(\mathcal{X}^{2})$ via the inverse transformation
of Eq.~(\ref{X^A}). The metric perturbations $O(\mathcal{X}^{2})$
in the last line of Eq.~(\ref{eq:a80}) should take specific forms
to satisfy the properties of the THZ coordinates as mentioned in Section~\ref{sec:DETERMINATION-OF-psiS}.
Our results show that \begin{eqnarray}
\tilde{g}_{00} & = & -1-\mathrm{\mathcal{E}}_{KL}\mathcal{X}^{K}\mathcal{X}^{L}-\frac{1}{3}\mathcal{E}_{KLM}\mathcal{X}^{K}\mathcal{X}^{L}\mathcal{X}^{M}+O(\rho^{4}/\mathcal{R}^{4}),\label{a68}\\
\tilde{g}_{0I} & = & \frac{{2}}{3}\epsilon_{IKP}\mathcal{B}^{P}{}_{L}\mathcal{X}^{K}\mathcal{X}^{L}-\frac{10}{21}\dot{\mathcal{E}}_{KL}\mathcal{X}^{K}\mathcal{X}^{L}\mathcal{X}_{I}+\frac{4}{21}\rho^{2}\dot{\mathcal{E}}_{KI}\mathcal{X}^{K}\nonumber \\
 &  & +\frac{1}{3}\epsilon_{IKP}\mathcal{B}^{P}{}_{LM}\mathcal{X}^{K}\mathcal{X}^{L}\mathcal{X}^{M}+O(\rho^{4}/\mathcal{R}^{4}),\label{a69}\\
\tilde{g}_{IJ} & = & \delta_{IJ}-\delta_{IJ}\mathrm{\mathcal{E}}_{KL}\mathcal{X}^{K}\mathcal{X}^{L}+\frac{5}{21}\epsilon_{IKP}\dot{\mathcal{B}}^{P}{}_{L}\mathcal{X}^{K}\mathcal{X}^{L}\mathcal{X}_{J}\nonumber \\
 &  & -\frac{1}{21}\rho^{2}\epsilon_{KPI}\dot{\mathcal{B}}_{J}{}^{P}\mathcal{X}^{K}-\frac{1}{3}\delta_{IJ}\mathcal{E}_{KLM}\mathcal{X}^{K}\mathcal{X}^{L}\mathcal{X}^{M}+O(\rho^{4}/\mathcal{R}^{4}),\label{a70}\end{eqnarray}
where $\rho^{2}=\mathcal{X}^{2}+\mathcal{Y}^{2}+\mathcal{Z}^{2}$
and indices $I,\, J,\, K,\, L,\, M,\, P=1,\,2,\,3$. The external
multipole moments are spatial, symmetric, tracefree tensors and are
related to the Riemann tensor evaluated on the particle's worldline
by\begin{eqnarray}
\mathrm{\mathcal{E}}_{IJ} & = & \left.\tilde{R}_{0I0J}\right|_{\mathrm{o}},\label{E_IJ}\\
\mathcal{B}_{IJ} & = & \frac{1}{2}\epsilon_{I}{}^{PQ}\left.\tilde{R}_{PQJ0}\right|_{\mathrm{o}},\label{B_IJ}\\
\mathcal{E}_{IJK} & = & \left[\nabla_{K}\left.\tilde{R}_{0I0J}\right|_{\mathrm{o}}\right]^{\mathrm{STF}},\label{E_IJK}\\
\mathcal{B}_{IJK} & = & \frac{3}{8}\left[\epsilon_{I}{}^{PQ}\nabla_{K}\left.\tilde{R}_{PQJ0}\right|_{\mathrm{o}}\right]^{\mathrm{STF}}.\label{B_IJK}\end{eqnarray}
where STF means to take the symmetric and tracefree part with respect
to the spatial indices $I,\, J,\, K$. The dot denotes differentiation
of the multipole moment with respect to $\mathcal{T}$ along the particle's
worldline. One should see that $\mathrm{\mathcal{E}}_{IJ}\sim\mathcal{B}_{IJ}\sim O(1/\mathcal{R}^{2})$
and $\mathcal{E}_{IJK}\sim\mathcal{B}_{IJK}\sim\dot{\mathcal{E}}_{IJ}\sim\dot{\mathcal{B}}_{IJ}\sim O(1/\mathcal{R}^{3})$
for consistency of the dimensions. The fact that all of the above
external multipole moments are tracefree follows from the assumption
that the background geometry is a vacuum solution of the Einstein
equations. These results agree with Eqs.~(17) and (18) of Paper I~\citet{detweiler-m-w(03)}
or Eqs.~(3.26a)-(3.26c) of Zhang~\citet{zhang(86)} to the lowest
a few orders.

\section{HYPERGEOMETRIC FUNCTIONS AND REPRESENTATIONS OF REGULARIZATION PARAMETERS\label{hyper}}

In Section \ref{sec:-REGULARIZATION-PARAMETERS} we define

\begin{equation}
\chi\equiv1-\alpha\sin^{2}\Phi\label{C1}\end{equation}
with

\begin{equation}
\alpha\equiv\frac{{J^{2}}}{r_{\mathrm{o}}^{2}+J^{2}}.\label{C2}\end{equation}
And we use\begin{eqnarray}
\left\langle \chi^{-p}\right\rangle \equiv\left\langle \left(1-\alpha\sin^{2}\Phi\right)^{-p}\right\rangle  & = & \frac{{2}}{\pi}\int_{0}^{\pi/2}\left(1-\alpha\sin^{2}\Phi\right)^{-p}d\Phi\nonumber \\
 & = & {}_{2}F_{1}\left(p,\frac{{1}}{2};1,\alpha\right)\equiv F_{p}.\label{C3}\end{eqnarray}
In particular, for the cases $p=\frac{{1}}{2}$ and $p=-\frac{{1}}{2}$
we have the following representations\begin{equation}
F_{1/2}={}_{2}F_{1}\left(\frac{{1}}{2},\frac{{1}}{2},1;\alpha\right)=\frac{{2}}{\pi}\hat{K}(\alpha)\label{C4}\end{equation}
and\begin{equation}
F_{-1/2}={}_{2}F_{1}\left(-\frac{{1}}{2},\frac{{1}}{2},1;\alpha\right)=\frac{{2}}{\pi}\hat{E}(\alpha),\label{C5}\end{equation}
where $\hat{K}(\alpha)$ and $\hat{E}(\alpha)$ are called complete
elliptic integrals of the first and second kinds, respectively.

If we take the derivative of $F_{1/2}$ with respect to $k\equiv\sqrt{\alpha}$
via Eq.~(\ref{C3}), we obtain\begin{equation}
\frac{{\partial F_{1/2}}}{\partial k}=-\frac{{F_{1/2}}}{k}+\frac{{F_{3/2}}}{k},\label{C6}\end{equation}
or using Eq.~(\ref{C4})\begin{equation}
\frac{{\partial\hat{K}}}{\partial k}=-\frac{{\hat{K}}}{k}+\frac{{\pi}}{2}\frac{{F_{3/2}}}{k}.\label{C7}\end{equation}
However, Ref.~\citet{arfken(01)} shows that \begin{equation}
\frac{{\partial\hat{K}}}{\partial k}=\frac{{\hat{E}}}{k\left(1-k^{2}\right)}-\frac{{\hat{K}}}{k}.\label{C8}\end{equation}
Thus, by comparing Eq.~(\ref{C7}) and Eq.~(\ref{C8}) we find the
representation \begin{equation}
F_{3/2}=\frac{{2}}{\pi}\frac{{\hat{E}}}{1-k^{2}}=\frac{{2}}{\pi}\frac{{\hat{E}}}{1-\alpha}.\label{C9}\end{equation}

Further, we can also find the representation for $F_{5/2}$. First,
taking the derivative of $F_{3/2}$ with respect to $k\equiv\sqrt{\alpha}$
via Eq.~(\ref{C3}) gives\begin{equation}
\frac{{\partial F_{3/2}}}{\partial k}=-\frac{{3F_{3/2}}}{k}+\frac{{3F_{5/2}}}{k}.\label{C10}\end{equation}
Also, using Eq.~(\ref{C9}) together with Eqs.~(\ref{C3})-(\ref{C5}),
another expression for the same derivative is obtained solely in terms
of complete elliptic integrals\begin{equation}
\frac{{\partial F_{3/2}}}{\partial k}=\frac{{2}}{\pi}\frac{{\left(1+k^{2}\right)\hat{E}-\left(1-k^{2}\right)\hat{K}}}{k\left(1-k^{2}\right)^{2}}.\label{C11}\end{equation}
Then, by Eqs.~(\ref{C9}), (\ref{C10}), and, (\ref{C11}) we find\begin{equation}
F_{5/2}=\frac{{2}}{3\pi}\left[\frac{{2\left(2-\alpha\right)\hat{E}}}{\left(1-\alpha\right)^{2}}-\frac{{\hat{K}}}{1-\alpha}\right].\label{C12}\end{equation}

Now, using Eqs.~(\ref{C4}), (\ref{C9}), and (\ref{C12}), we may
rewrite the non-zero \emph{B}-terms of regularization parameters,
Eqs. (\ref{eq:13})-(\ref{eq:15}) in Section \ref{sec:MODE-SUM-DECOMPOSITION-AND}
as \begin{equation}
B_{t}=\frac{{q^{2}}}{r_{\mathrm{o}}^{2}}\frac{{E\dot{{r}}\left[\hat{K}(\alpha)-2\hat{E}(\alpha)\right]}}{\pi\left(1+J^{2}/r_{\mathrm{o}}^{2}\right)^{3/2}},\label{C13}\end{equation}

\begin{equation}
B_{r}=\frac{{q^{2}}}{r_{\mathrm{o}}^{2}}\frac{{\left(\dot{r}^{2}-2E^{2}\right)\hat{K}(\alpha)+\left(\dot{r}^{2}+E^{2}\right)\hat{E}(\alpha)}}{\pi\left(1-2M/r_{\mathrm{o}}\right)\left(1+J^{2}/r_{\mathrm{o}}^{2}\right)},\label{C14}\end{equation}

\begin{equation}
B_{\phi}=\frac{{q^{2}}}{r_{\mathrm{o}}}\frac{{\dot{r}\left[\hat{K}(\alpha)-\hat{E}(\alpha)\right]}}{\left(J/r_{\mathrm{o}}\right)\left(1+J^{2}/r_{\mathrm{o}}^{2}\right)^{1/2}},\label{C15}\end{equation}
  which are exactly the same to the results of Barack and Ori \citet{barack-ori(02)}.


\begin{thebibliography}{10}
\bibitem[1]{detweiler-whiting(03)}S. Detweiler and B. F. Whiting,
Phys. Rev. D \textbf{67}, 024025 (2003), gr-qc/0202086.

\bibitem[2]{dirac(38)}P. A. M. Dirac, Proc. R. Soc. (London) \textbf{A167},
148 (1938).

\bibitem[3]{dewitt-brehme(60)}B. S. DeWitt and R. W. Brehme, Ann.
Phys. \textbf{9}, 220 (1960).

\bibitem[4]{mino-sasaki-tanaka(97)}Y. Mino, M. Sasaki, and T. Tanaka,
Phys. Rev. D \textbf{55}, 3457 (1997).

\bibitem[5]{quinn-wald(97)}T. C. Quinn and R. M. Wald, Phys. Rev.
D \textbf{56}, 3381 (1997).

\bibitem[6]{quinn(00)}T. C. Quinn, Phys. Rev. D \textbf{62}, 064029
(2000).

\bibitem[7]{detweiler-m-w(03)}S. Detweiler, E. Messaritaki, and B.
F. Whiting, Phys. Rev. D \textbf{67}, 104016 (2003).

\bibitem[8]{thorne-hartle(85)}K. S. Thorne and J. B. Hartle, Phys.
Rev. D \textbf{31}, 1815 (1985).

\bibitem[9]{zhang(86)}X.-H. Zhang, Phys. Rev. D \textbf{34}, 991
(1986).

\bibitem[10]{barack-ori(02)}L. Barack and A. Ori, Phys. Rev. D \textbf{66},
084022 (2002), gr-qc/0204093.

\bibitem[11]{mino-nakano-sasaki(02)}Y. Mino, H. Nakano, and M. Sasaki,
Prog. Theor. Phys. \textbf{108}, 1039 (2002), gr-qc/0111074.

\bibitem[12]{barack-ori(00)}L. Barack and A. Ori, Phys. Rev. D \textbf{61},
061502(R) (2000).

\bibitem[13]{weinberg(72)}S. Weinberg, \emph{Gravitation and Cosmology}
(Wiley, New York, 1972).

\bibitem[14]{jackson(99)}J. D. Jackson, \emph{Classical Electrodynamics}
(Wiley, New York, 1999), 3rd ed.

\bibitem[15]{mathews-walker(70)}J. Mathews and R. L. Walker, \emph{Mathematical
Methods of Physics} (W. A. Benjamin, New York, 1970), 2nd ed.

\bibitem[16]{arfken(01)}G. B. Arfken and H. J. Weber, \emph{Mathematical
Methods for Physicists} (Harcourt/Academic Press, San Diego, 2001),
5th ed.

\bibitem[17]{kim-detweiler(04)}D.-H. Kim and S. Detweiler, {}``Regularization
parameters for gravitational self-force calculations in the Schwarzschild
geometry'' (in preparation)
\end{thebibliography}
\end{document}